\documentclass[mathleft
]{an}
\usepackage{graphicx}
\usepackage{rotating}
\usepackage{lscape}
\usepackage{longtable}
\usepackage{times}
\begin{document}

\sloppy

\Pagespan{789}{}
\Yearpublication{2011}%
\Yearsubmission{2010}%
\Month{11}%
\Volume{999}%
\Issue{88}%


\title{Solar activity around AD 775 from aurorae and radiocarbon}

\author{R. Neuh\"auser\inst{1} \thanks{Corresponding author: \email{rne@astro.uni-jena.de}}
\and D.L. Neuh\"auser\inst{2}
}

\titlerunning{Solar activity around AD 775}
\authorrunning{Neuh\"auser \& Neuh\"auser}

\institute{
Astrophysikalisches Institut und Universit\"ats-Sternwarte, FSU Jena,
Schillerg\"a\ss chen 2-3, 07745 Jena, Germany
\and
Schillbachstra\ss e 42, 07743 Jena, Germany
}

\received{Nov 2014}
\accepted{27 Feb 2015}
\publonline{ }



\abstract{A large variation in $^{14}$C around AD 775 
has been considered to be caused by one or more solar super-flares within one year.
We critically review all known aurora reports 
from Europe as well as the Near, Middle, and Far East 
from AD 731 to 825 and find 39 likely 
true aurorae plus four more potential aurorae
and 24 other reports 
about halos, meteors, thunderstorms etc., 
which were previously misinterpreted as aurorae or misdated;
we assign probabilities for all events according to five aurora criteria.
We find very likely true aurorae in AD 743, 745, 762, 765, 772, 773, 793, 796, 807, and 817.
There were two aurorae in the early 770s observed near Amida 
(now Diyarbak\i r in Turkey near the Turkish-Syrian border),
which were not only red, but also green-yellow --
being at a relatively low geomagnetic latidude, 
they indicate a relatively strong solar storm. 
However, it cannot be argued that those aurorae 
(geomagnetical latitude $43$ to $50^{\circ}$,
considering five different reconstructions of the geomagnetic pole)
could be connected to one or more
solar super-flares causing the $^{14}$C increase around AD 775: 
There are several reports about low- to mid-latitude aurorae 
at $32$ to $44^{\circ}$ geomagnetical latitude in China and Iraq;
some of them were likely observed (quasi-)simultaneously in two of
three areas (Europe, Byzantium/Arabia, East Asia),
one lasted several nights, and some indicate a particulary strong geomagnetic
storm (red colour and dynamics),
namely in AD 745, 762, 793, 807, and 817
-- always without $^{14}$C peaks.
We use 39 likely true aurorae as well as historic reports about sunspots 
together with the radiocarbon content from tree rings
to reconstruct solar activity: 
From AD $\sim 733$ to $\sim 823$, 
we see at least nine Schwabe cycles;
instead of one of those cycles, there could be two short, weak cycles --
reflecting the rapid increase to a high $^{14}$C level since AD 775,
which lies at the end of a strong cycle.
In order to show the end of the dearth of naked-eye sunspots, 
we discuss two more Schwabe cycles until AD $\sim 844$.
The $^{14}$C record (from both Intcal and Miyake et al. 2013a) is anti-correlated
to auroral and sunspot activity, 
as expected from solar wind modulation 
of cosmic rays which produce the radiocarbon. 
}

\maketitle

\section{Introduction: The $^{14}$C variation AD 770s}

Miyake et al. (2012, henceforth M12) found a variation in the $^{14}$C to $^{12}$C isotope ratio
in two Japanese trees around the year AD 775; 
they excluded supernovae as a possible cause due to the lack
of any historic observations and of any young nearby supernova remnants; they also excluded
solar super-flares as cause, because they would not be hard enough to explain the
$^{14}$C to $^{10}$Be production ratio observed (M12).
Usoskin \& Kovaltsov (2012) and Melott \& Thomas (2012), 
then also Thomas et al. (2013) and Usoskin et al. (2013, henceforth U13)
suggested that a solar super-flare with $\ge 24^{\circ}$ beam size could have
caused the event, in particular if 4 to 6 times less $^{14}$C was
produced than calculated in M12 due to a different carbon cycle model.
In that case, a solar super-flare would be required which 
would be only a few times larger than the Carrington event in AD 1859 (U13).
However, partly because neither the Carrington event (presumably strongest solar flare in the 19th century)
nor the solar proton event in 1956 (hardest and strongest solar flare in the 20th century) 
are detected as $^{10}$Be, $^{14}$C (problematic since the 1950s due to atomic bombs), 
nor cosmic rays (already available in 1956) increases, 
Cliver et al. (2014) and Neuh\"auser \& Hambaryan (2014)
doubt the solar super-flare hypothesis. 
If the event was due to one or several solar super-flares,
strong aurorae should have been observed even at geomagnetic latitudes
lower than during the Carrington event 
(e.g., Hawai$^{\prime}$i (Fritz 1873) with a geomagnetic latitude of $\sim 20^{\circ}$).
Hambaryan \& Neuh\"auser (2013) suggested that a short Gamma-Ray-Burst could have caused the event,
because all observables including the $^{14}$C to $^{10}$Be production ratio are consistent with such a burst;
Pavlov et al. (2013) confirmed those rough estimates with more precise calculations.

To evaluate whether a solar super-flare might have been responsible for
the $^{14}$C increase around AD 775, 
one can search for exceptionally strong aurorae and, more generally, 
one has to consider the status of the solar activity at that time, 
including the phase of the Schwabe cycle.

Solar activity undergoes a well-known Schwabe cycle
with maxima and minima every $\sim 10$ yr according to sunspot observations (Schwabe 1843), 
or $131 \pm 14$ months (Hathaway \& Wilson 2004, Hathaway 2010),
or $10.4 \pm 1.2$ yr (Usoskin et al. 2001), the average since AD 1750.
The Schwabe cycle length since AD 1750 (from minimum to minimum)
varied between $\sim 8$ and $\sim 14$ yr (Hoyt \& Schatten 1998, Hathaway 2010);
as an exception, it may have been as low as 5-7 yr,
if there was an additional short weak cycle inside cycle 
no. 4\footnote{According to Usoskin et al. (2001), the extra minimum between the two cycles inside
cycle no. 4 was in AD 1793.1, and the following next minimum was
either 1799.8 (Usoskin et al. 2001) or 1798.3 (Hoyt \& Schatten 1998, Hathaway 2010),
resulting in a cycle length of only 5.2 to 6.7 yr. 
While Usoskin et al. (2009) could show that the butterfly diagram reconstructed
for that time with newly found old sunspot drawings (Arlt 2009) is not inconsistent with the hypothesis to split the
cycle into one normal and one extra short weak cycle,
Zolotova \& Ponyavin (2011) discussed whether the unusual shape and length of the long cycle no. 4 could
be explained just by a {\em phase catastrophe}; most recently, Karoff et al. (2015) studied the $^{10}$Be data
around that time: By taking into account the hysteresis effect (different time dependence of cosmic ray input into the solar
system in A+/A- cycles), they reconstructed the solar modulation potential and the sunspot (group) numbers around AD 1800 
from $^{10}$Be for both cases, with and without splitting cycle no. 4 into two cycles (4a, 4b)
and considering their polarity (A+ or A-), and then compared the results with real sunspot data;
they concluded that, after splitting cycle 4 into 4a and 4b, 
the $^{10}$Be-calculated solar modulation potential would be better
consistent with the sunspot data -- even though 
some of their modulation-potential-maxima do not coincide with sunspot maxima.} 
or as high as 17-18 yr (maximum of cycle no. 4 to maximum of cycle no. 5), 
if cycle no. 4 remains undivided (Hathaway 2010).

Solar activity can be reconstructed for past centuries from sunspot and aurora observations,
which may, however, be incomplete, biased, and inhomogeneous; hence, radio\-nuc\-lide archives on Earth
may yield more complete and homogeneous results:
The larger the solar activity, the larger the number of sunspots and aurorae,
and the stronger the solar wind and, hence, the smaller the incoming cosmic-rays, 
hence, a {\em decrease} in the
production ratio of radionuclides (review by Usoskin 2013).
Both the occurrence of aurorae and the $^{14}$C production depend on the geomagnetic field,  
which did not change significantly during the century studied here
(see e.g. Korte \& Constable 2011 and Korte et al. 2011),
so that we can neglect these dependences here.
$^{14}$C incorporation in tree rings is also related to the carbon cycle
and cannot give a time resolution of better than 1 yr 
($^{10}$Be has lower accuracy in absolute timing),
while sunspot and aurora observations can indicate solar activity changes on time-scales of days.
Below, we will consider the geomagnetic latitudes of aurorae around AD 775.

We use here aurora and sunspot observations as well as $^{14}$C
to specify phases of maximal and minimal solar activity;
our results revise previous reconstructions of the Schwa\-be cycle for the 8th century,
which partly misinterpreted other phenomena as auorae,
partly misdated some likely true aurorae, 
and/or which did not use all known aurora reports
(Schove 1955, U13);
furthermore, Schove (1955) assumed that there were always nine Schwabe cycles per century
with a length of 8-16 yr.
We investigate, whether there were any particularly strong aurora observations,
which may be connected to one or more giant solar super-flare(s).

We restrict this study to the time from AD 731 to AD 825, i.e. around AD 775.
An even broader study from AD 550 to 845 (including the Dark Age Grand Minimum) 
will present an aurora catalog with most of the original texts and
additional information on the historic sources (in prep.).
We show in this paper already the clusters of aurorae around AD 829 and 839.

In Sect. 2, we discuss on the one and only sunspot observation in our study period (Sect. 2.1); 
in addition, we list the reports in an extended period until AD 844,
in order to show the end of the dearth of (naked-eye) sunspots (Sect. 2.2);
in Sect. 2.3, we discuss the sample completeness.
In Sect. 3, we first explain the criteria for constructing our aurora catalog (Sect. 3.1)
and list all likely true aurorae from AD 731 to 825 (Sect. 3.2),
then we discuss the question of strong aurorae around AD 775 (Sect. 3.3),
list further events which were previously misinterpreted and/or misdated as aurorae (Sect. 3.4),
and also compare our catalog with other previously published global aurora catalogs (Sect. 3.5).
Finally, in Sect. 4, we reconstruct the Schwabe cycles with aurorae, sunspots, and
radiocarbon for the period studied, discuss limitations, and conclude with our results.

\section{Sunspots from AD 731 to 825 (844)}

There are no sunspot records known for the 7th and 8th centuries AD (AD 580-806),
see e.g. Keimatsu (1973), Clarke \& Steph\-enson (1978), Wittmann \& Xu (1987), 
Yau \& Stephenson (1988), Xu et al. (2000), Vaquero et al. (2002), or Vaquero (2007).

\subsection{AD 731 -- 825}

For the period studied here (AD 731 to 825), 
the only known report about a sunspot is dated to AD 807 March (for Europe),
for which we quote also the context (text in round brackets are significant variants
from the translation by Scholz \& Rogers (1970), text in square brackets are Newton's and
our additions]: \\
{\em (The Emporer celebrated Christmas at Aachen.) 
Again on the 4th calends March} [AD 807 Feb 26] {\em was an eclipse
of the moon, and flames appeared that night of an amazing brightness}
[Lat.: {\em acies eadem nocte mirae magnitudinis}]
{\em (and in the same night enourmous battle lines appeared in the sky), 
and the Sun stood in the 11th degree of Pisces and the moon in the 11th degree of Virgo.
And, still more, the star Mercury on the 16th calends April} [AD 807 March 17]
{\em was seen in the Sun like a small black spot} 
[Lat.: {\em in sole quasi parva macula, nigra tamen}], 
{\em a little above the center of that very body,
and it was seen by us for 8 days, but when it first entered, and when it left, clouds
kept us from observing (we could not observe it well because of clouds).} \\
citing from Newton (1972) for report {\em Annales Loiselianos}, a variant (copied about AD 814 by a person
named Loisel) of the Annales Laurissenes maiores from the monastery of Lorsch in Germany (Newton 1972), 
later called the Royal Frankish Annals (a similar, but shorter report is also found
in Einhard's biography of Charlemagne); they were written by different authors and cover the
period from AD 741 to 829, the last four decades written by contemporaneous eyewittnesses;
see Rau (1955), Newton (1972), or McKitterick (2008) 
for the different manuscripts, etc. 
Since there was no Mercury transit in that year and since it cannot last several days,
this sighting was interpreted as first European record for a sunspot (e.g. Newton 1972).

The report for that year also mentions several eclipses, all with the correct dates --
considering that the dates are given for the system at that time (9th century), 
namely that the preceding night may have belonged to what we now regard as next day;
the lunar eclipse AD 807 Feb 26 centered around 3:49h CET started on the evening of AD Feb 25,
so that the date Feb 26 is given; hence, the whole report appears highly credible.
The {\em flames} reported for AD 807 Feb 26 were probably aurorae during the lunar
eclipse (Newton 1972).

It is, however, quite anachronistic for a 9th century report from central Europe
to notice a sunspot or to report a {\em Mercury transit}.
For several centuries before and after that time, there is no other similar report from
Europe about transits or sunspots.
This does not mean that the report is misdated or fabricated.

In the Royal Frankish Annals, the source of this detailed report,
the next event narrated after the above quotation is the visit of
a diplomatic delegation from the Arab (given as {\em Persia}) 
caliphate in Baghdad (citing from Scholz \& Rogers 1970): \\
{\em Radbert, the Emporer's emissary, died on his way back from the East.
The envoy of the king of Persia by the name of $^{c}$Abdall\={a}h} 
[H\={a}r\={u}n ibn Mu\d{h}ammad ibn $^{c}$Abdall\={a}h, also called called {\em A-lun} in the Chinese Tang Annals,
best known as H\={a}r\={u}n al-Rash\textit{\={\i}}d, caliph until AD 809]
{\em came to the Emporer with
monks from Jerusalem, who formed an embassy from the patriarch Thomas. Their names
were George and Felix. This George is abbot of Mount Olives, a native German and
called, by his real name, Egilbald ...} \\
Then, gifts are mentioned including a complicated water clock that struck the hours
(as used since centuries in China) indicating technology transfer.
It is possible that knowledge transfer in astronomy also happened,
namely that the Arab delegation (also in contact with China) 
told and/or taught the court of Charlemagne about those astronomical observations
including the eclipses and the presumable Mercury transit --  
Arabic scholars were interested in transits since at least AD 840 
(probably even earlier), see below. The superior astronomical knowledge of that
delegation may have motivated Charlemagne to call for an empire-wide astronomical congress,
which took place a few years later.

The AD 807 report assumes that the observation was caused by a Mercury transit; 
however, as we know today, such a transit does not last for several days, 
and there also was no such transit in AD 807 March;
the closest real Mercury transit happened on AD 806 Apr 27\footnote{www.projectpluto.com/transits.htm$\#$mercury}, 
the closest real Venus transit was on AD 797 May 22\footnote{calculated by Fred Espanak, 
NASA/GSFC, eclipse.gsfc.nasa.gov/ transit/catalog/};
Venus was at large western elongation and Mercury at large eastern elongation AD 807 Mar 17 to 24.
It is surprising that the sighting was interpreted as a Mercury transit, because Mercury
was well visible on the evening of AD 807 Mar 17-24 (at $\ge 10^{\circ}$ above horizon
at Aachen, Germany, the location of Charlemagne's court) being around 1st to 2nd magnitude.

According to all three major variants of the solar system architecture (geocentric following Aristotle,
heliocentric following Aristarch of Samos and much later Copernicus, 
or intermediate geo-helio-centric) 
Mercury and Venus were expected to be able to transit the Sun.
However, the dimension of and distances within the solar system were not known well,
so that, e.g., the occurence time and duration of transits were not known;
hence, Arabic and European medieval scholars (after Greek papers were 
translated first to Arabic starting in the 8th century, then to Latin) 
tried to observe such transits to learn about the solar system architecture;
see e.g. Goldstein (1969) for a discussion.
If the different non-zero inclinations of the planetary orbits are not known,
one would expect a transit at each inferior conjunction of Venus or Mercury;
and it may have been unknown to those scholars 
whether an upcoming conjunction was inferior or superior.
Following Aristotle, the Sun was assumed to be a perfect sphere
without inhomogenities, so that spots were not considered possible.
Chinese observers of the last millenia interpreted similar observations as spots 
or some kind of vapour on (or before) the Sun.

\subsection{AD 826 -- 844}

After the long dearth of (naked-eye) sunspots until AD 807,
there were several reports in the coming decades;
we list here those until AD 844, which will be helpful later
for Schwabe cycle reconstructions.
 
There was a long-duration sunspot sighting (three solar rotation periods) later reported by Ibn al-Qif\d{t}i in Arabic
for AD 840: \\
{\em Ghars al-Ni$^{c}$ma Mu\d{h}ammad ibn al-Ra$^{\prime}$\textit{\={\i}}s Hil\={a}l ibn al-Mu\d{h}assin al-\d{S}\={a}bi$^{\prime}$ said in his book:
I found in the handwritings of Ja$^{c}$far b. al-Muktaf\textit{\={\i}} ... that in the year 225} 
[lunar years since AD 622 July 16 or 15, the start of the Islamic year in which the Hijra took place]
{\em during the caliphate of al-Mu$^{c}$ta\d{s}im there appeared a black spot} 
[Arabic: {\em nukta sawd\={a}$^{\prime}$}] {\em close to the middle
of the Sun. This took place on Tuesday, 19 Rajab 225} [AD 840 May 25], {\em and when two days had
gone from this date, i.e. after 21 Rajab, events occured. Al-Kind\textit{\={\i}} mentioned that this spot
lingered on the Sun for 91 days} [i.e. until AD 840 Aug 23] {\em and soon thereafter Mu$^{c}$ta\d{s}im died} 
[AD 842] ... {\em Al-Kind\textit{\={\i}} mentioned that this spot was due to the occulting of the Sun by Venus, 
and their clinging together for this period ...} \\
citing from Goldstein (1969). \\
The first part of the text brings a clear observation, {\em a black spot close to the middle
of the Sun}, for which the observing date is given; this observation was 
seen as a negative portent: {\em when two days had gone from this date events occured}. \\
Then, an additional report from the astronomer al-Kind\textit{\={\i}} is added, 
according to which the spot was a Venus transit that would have lasted {\em for 91 days}.\\
In the middle of July 840, Venus was in conjunction with the Sun, 
and Mercury also in early June and August 840, so that the Arab scholars may have expected 
either a Mercury transit around the days of first and last appearance of the {\em spot},
and/or a Venus transit during the period.  \\
However, it is surprising that an astronomer could come to such an interpretation:
Venus was at largest elongation in AD 840 May,
and it was also well visible in the evening sky in August 840 being brighter than -3 mag.
The closest real Venus transits were
on AD 797 May 22 and AD 910 Nov 23\footnote{footnote 2},
the closest Mercury transits were on AD 839 Apr 25 and AD 842 Oct 26/27\footnote{footnote 1}.
Given that a single spot cannot be observed for 91 days, we have to assume
that different spots (or groups) were observed in this period or at least on the first and last dates given;
the above quotation from Ibn al-Qifti is probably composed of two sources, one from al-\d{S}\={a}bi$^{\prime}$
and one from Al-Kind\textit{\={\i}}; we plot two spots in Fig 1.
The location of the spot ({\em close to the middle of the Sun}) was given only
for the first observing day. 

Chinese observers noticed several more spots: 
\begin{itemize}
\item {\em Within the Sun there was a black vapour like a cup} observed
at the Tang capital (today Xian) in the Shaanxi province on AD 826 May 7 
(Keimatsu 1974, Wittmann \& Xu 1987, Xu et al. 2000) from Xin Tang shu, Tianwen zhi,
a {\em cup} may refer to a spot group.  
\item {\em Within the Sun there was a black spot} observed
in Xian on AD 826 May 24 
(Keimatsu 1974, Wittmann \& Xu 1987, Xu et al. 2000) from Xin Tang shu, Tianwen zhi.  
\item {\em There was a black, weird thing in the Sun, and it looked like struggeling against it} observed
in Xian on AD 829 Jan 20 (Keimatsu 1974) from T'ang-shu, possibly also a group, since it is described as {\em weird},
hence not circular. 
\item {\em There was a black, weird thing in the Sun, and it looked like struggeling against it} observed
in Xian on AD 832 (one or several days) between Apr 5 and 20 (Keimatsu 1974) from T'ang-shu; since the texts
of this and the previous report are identical except the dates, one of the two might be a misdated duplication,
but both are from T'ang-shu. 
\item {\em Within the Sun there was a black spot} observed in Xian on AD 832 Apr 21 
(Keimatsu 1974, Wittmann \& Xu 1987, Xu et al. 2000) from Xin Tang shu, Tianwen zhi. 
\item {\em A black vapour covered the Sun} (in T'ang-shu), {\em a black vapour ground the Sun} (in Wen-ksien-t'ung-k'ao)
observed in Xian on AD 832 May 6 (Keimatsu 1974, Clark \& Stephenson 1978, Yau \& Stephenson 1988,
Xu et al. 2000) from Xin Tang shu, Tianwen zhi -- maybe again a group, since the reports do not
mention a simple {\em spot}. 
\item {\em Within the Sun there was a black spot as large as a hen's egg. The Sun was red like ochre and in the 
daytime it was like evening until day 20} (Dec 24) observed in Xian on AD 837 Dec 22-24 
(Keimatsu 1974, Wittmann \& Xu 1987, Xu et al. 2000) from Xin Tang shu, Tianwen zhi;
the latter effect does not indicate a noticable effect of the large spot (group) on the day time brightness
of the Sun, but could have been caused by large amounts of dust in the air -- e.g. due to a volanic eruption
as it was argued for another such Chinese naked-eye sunspot observation (Schove 1951) -- also 
facilitating the naked-eye observation of the sunspot.
\item {\em In the Sun, there was a black spot} 
from Keimatsu (1974) and Xu et al. (2000) from Xin Tang shu, Tianwen zhi (AD 841 Dec 30).
\end{itemize}
We plot all those Chinese, Arabic, and European naked-eye sunspot reports in Fig. 1.

\subsection{Completeness}

The fact that the Arabic and Chinese spot observations listed above do not overlap,
but are in fact disjunct, shows that the historic observations and/or records 
are most certainly not complete, but complimentary. The fact that there is not one single
spot observed in two continents indicates that the reason for the reports being disjunct
cannot simply be bad weather at one location, but it indicates that the observers did not
always looked for spots (as they were not considered very important). 
The Arabs may have observed the Sun only when they expected a transit.
On the other hand, as Ogurtsov et al. (2002) wrote:
{\em The reliability of the information on solar activity, extracted from historical chronicles,
was analysed in many works ... and it was shown that the Oriental historic data really reflect
such features of sunspot activity as decadel and century-scale cycles, butterfly diagrams,
deep Maunder-type minima} (Ogurtsov et al. 2002).

Given that the Chinese Tang Dynasty capital Chang-an (now Xian)
was occupied due to the Lu-Shan rebellion about AD 755-760,
it is possible that observations were interrupted and/or that original documents 
(of this or earlier periods) were lost (e.g. Yau \& Stephenson 1988).
There are both day- and night-time observations by Chinese in the period studied here
including solar eclipses and halos (Clark \& Stephenson 1978, Yau 1988, Xu et al. 2000).
Furthermore, as it was argued by Clark \& Stephenson (1978), it is quite likely that
Chinese day-time observations were actually carried out throughout the 7th and 8th centuries.
The reason for the lack of sunspot records for the entire 7th and 8th centuries is not yet known;
even political and ideological reasons cannot be excluded.

The Chinese astronomers did report occultations of stars by planets for 
AD 761 Feb 6, AD 773 May 4 (both Jupiter occulting $\beta$ Sco), for AD 773 Nov 30 (Venus occulting $\beta$ Sco),
and then also for AD 780 Dec 1 (Jupiter occulting the Praesepe cluster), 
and then also grazing occultations or very close conjunctions
between planets and stars  
for AD 767 Oct 17/18 (Mars and $\lambda$ Sgr), 768 Oct 23 (Venus and $\eta$ Vir), 768 Nov 3 (Mars and $\eta$ Vir),
and 768 Nov 4 (Venus and $\theta$ Vir) --
all those occultations and conjunction were confirmed by modern calculations (Hilton et al. 1988).
Another occultation is reported below (Sect. 3.4), an occultation of the Pleiades by the moon on AD 761 Dec 13,
also confirmed by modern calculation.

Furthermore, a comet was observed in AD 773 Jan in China and in Japan (Chapman et al. 2014),
and a meteor shower was observed on AD 773 Jan 10 in Japan (Imoto \& Hasegawa 1958), probably from the Quadrantids,
one of the most rich showers.\footnote{Meteor
showers were also observed, e.g., on AD 745 Jan 1 or at least 
early January ({\em at the beginning of Kanon II (Jan), bolides appeared from all 
sides running through the air ... In the middle of the sky, 
it was seen during the night something like a large column of fire},
i.e. probably both meteors and an aurora, as reported by Michel le Syria, 
see Sect. 3.2 below), which indicates that this meteor shower had
appeared already in AD 745 and 773, so that it cannot have formed 
in the Ching-yang event in AD 1490 March/April as was claimed by Jenniskens (2004).}
A guest star was observed in Korea in AD 776, 
namely the nova or supernova candidate listed as {\em Hye Sung} (an ancient Korean comet name) in Chu (1968) for 
AD 776 Jun 1-30 in Tau-Aur (from the Korean Lee Dynasty chronology during the reign of He Gong),
which is more likely a comet (observed for 30 days, while a supernova is visible for several months)
and named with an old Korean comet name, possibly because it was moving relative to the stars,
possibly without a tail or with a very small tail.
Another comet dated to AD 776 Jan (e.g. Kronk 1999) was unintentionally copied and misdated,
but was in fact in AD 767 Jan (Chapman et al., to be submiited).
All these records show that night-time observations were performed in Eastern Asia in the relevant epoch.

\section{Aurorae from AD 731 to 825}

Aurorae indicate a disturbance of geomagnetic activity due to solar magnetic activity
(see, e.g., Vaquero \& Vazquez (2009) and references therein): 
Transient shock activity (the most well-known aurorae) is related to solar flares,
coronal mass ejections, and reconnections, so they peak around the 
sunspot maximum ($\sim 1/3$ of all aurorae).
Coronal hole aurorae due to the long-lived high-speed wind 
are considered to be the most frequent type of aurorae ($\sim 1/2$ of all aurorae);
severe geomagnetic disturbances and, hence, aurorae occur when polar
and equatorial coronal holes merge --
leading to aurora peaks 1-3 yr before and after the sunspot maximum
plus additional enhanced geomagnetic activity in the declining phase.

For the period studied here, 
in all of the three major areas from where written documents are available
(Europe, Byzantium/Arabia, and East Asia), the reports about aurorae (and sunspots) 
may be incomplete (see also Sect. 2.3): \\
China: Occupation of the Tang Dynasty capital Chang-an (now Xian)
during the Lu-Shan rebellion (AD 755-760) as mentioned above. \\
Japan: Relocation of the capital and imperial court
from Nara to Heian-kyo (now Kyoto), far from Buddhist influence,
in AD 784 and 794, respectively. \\
Byzantium: The 1st and 2nd Iconoclasm lasted from AD 726-787 and AD 814-842
with much internal turmoil and external wars including attacks by muslim Arabs. \\
Arabia: Non-peaceful transition from Umayyad to Abbasid dynasty in AD 750,
then foundation of new capital Baghdad in AD 762;
Arabic scholars had started to translate ancient (Greek, Persian, and Indian) texts 
about astronomy and other science not earlier than in the middle of the 8th century,
and then started with a first astronomical observatory in the first quarter of the 9th century
(the aurora presented here for the first time for AD 793 from Iraq is the earliest dated
Arabic aurora known so far). \\
Continental Europe: After the fall of the Western Roman Empire in the 5th/6th century, 
science and documentation of natural events were at a minimum throughout 
most of Europe (with few exceptions like, e.g.,
Gregor of Tours, who died in AD 594); almost 200 yr later,
the large European/Frankish empire by Charlemagne was unified (by him after the sudden death of his brother)
in AD 771; Italy was occupied in AD 774, so that a reception of Roman culture began; 
the Royal Frankish Annals cover AD 741-829, but the reports for the first few decades are very brief 
(based on earlier Easter tables), and only the last four decades were written by contemperous 
eyewitt\-nesses. \\
The Irish and Scotish chronicles as well as the (later compiled)
Anglo-Saxon Chronicle do report aurorae in the 8th and 9th centuries,
but they are all relatively far north, i.e. not neccessarily indicating strong solar storms. 

Given the constant, but rather low rate of reports 
(about true aurorae and other celestial events, see Sect. 3.2 and 3.4), 
there is no evidence for extended periods where reports would be completely missing or lost
(as also found in Sect. 2.3). 

\subsection{Aurora criteria and sources}

Previous compilations of historic aurorae include other phenomena misinterpreted as aurorae,
such as halo displays, meteors, or weather phenomena, we therefore have to review all previous reports
and compilations carefully.

For a critical investigation of all those aurorae, we consulted most of the original texts
in medieval Latin, Greek, English, Irish, and Danish,
we have also reviewed scholarly translations of Arabic, Syriac, and Chinese texts,
and translated -- some of them for the first time -- to modern English 
(sometimes using German or French translations).
We should keep in mind that the different chronicles are mostly copies of copies etc. of the
original (but lost) hand-writings, so that sometimes the text and/or the dating can be somewhat uncertain.
Only very few of the texts are autographs, whose authors or informants are true eyewitnesses. 
We used the source or copy closest in time.
Note that different definitions of the year were used. \\
In all reports, the appearance of the sighting is described, not its physical nature.
While the far Eastern texts are often written by professional astronomers with systematic
wording, those from Europe and Byzantium are inhomogeneous. 
Reports from Europe and Near East about celestial phenomena mention {\em signs}
like eclipses, comets, meteors, aurorae, and halos; in Christian reports
in the 8th and 9th centuries, words like {\em halo} or {\em aurora} are never used:
E.g., a phenomenon described as {\em red dragon on sky} or similar
could be either an auroral display or a halo arc effect;
often, it is neither mentioned whether or not it happened at night nor in which direction;
also, the interpretation of the {\em sign} in an astrological sense as portent for events and/or the
context (often given) does not always lead to a clear solution
(see Neuh\"auser \& Neuh\"auser 2014).

Most of the known oriental aurora records are found in Matsushita (1956), 
Keimatsu (1973, 1974),\footnote{The summary
paper written by N. Fukushima after M. Keimatsu had passed away on 
1976 Jul 3 is cited by us as Keimatsu \& Fukushima (1976); 
it is sometimes cited as Keimatsu (1976).
From Matsushita (1956) and Keimatsu (1973, 1974), 
we include all those events (as aurorae or misinterpreted events) which were classified by them with
(a high) probability 1-3 (out of five); in Keimatsu (1973, 1974), this meant {\em certain} for 1,
{\em very probable} for 2, and {\em probable} for 3; events classified by
Keimatsu (1973, 1974) with lower probability ({\em doubtful} or {\em unlikely}) are discussed 
in this paper, when they were classified as true aurorae by others, 
like Yau et al. (1995), Xu et al. (2000), or Usoskin et al. (2013);
in addition, we did check all events classified in Keimatsu (1973, 1974) 
as {\em probable to doubtful} (3-4) or lower and included two such events
here (AD 804 and 819), because they appear as (very) possible aurorae.
The remaining events from Keimatsu (1973, 1974) ({\em probable to doubtful}, {\em doubtful} or {\em unlikely})
in our study period were in the following years: AD 748, 756, 757, 760, 764, 765, 787, 798/799, 
808, 811, 814, and 824; these reports are mainly about meteors and bolides;
for completeness, we will cite them in Neuh\"auser \& Neuh\"auser (in prep.).}
Dai \& Chen (1980), Yau et al. (1995), or Xu et al. (2000); for the Kei\-mat\-su (1973, 1974) reports,
we cite new translations in Yau et al. (1995) or Xu et al. (2000), when available.

Most known occidental records are found 
in de Mairan (1733), Sch\"oning (1760), Jeremiah (1870), Fritz (1873), 
Schove (1955, 1964, 1984), Link (1962), Newton (1972), and Dall'Olmo (1979), 
plus a few also in Bone (1996), McCarthy \& Breen (1997), and U13;
Dall'Olmo (1979), 
Rada \& al-Najeh (1997),
Cook (2001),
and Basurah (2005, 2006) cite also a few detailed reports from the Near and Middle-East.

There are also two world-wide compilations, one just with aurora reports (Silverman 1998),
and one which also includes other astronomical events (Hetherington 1996),
both of them copying from other catalogs, but both with some shortcomings, e.g. a {\em red cross} seen
in England (in AD 776, see Sect. 3.4) is mis-interpreted by both of them as aurora and listed for AD 773, 774, and 776,
partly quoting secondary and misleading literature.

Since there are many events listed in published aurora catalogs which are not aurorae, but some other effect
(like e.g. halos, eclipses, meteors, rainbows, volcano eruptions, comets, novae, weather phenomena, etc.), 
we establish five criteria for the {\em likeliness} of the event to be an aurora,
which are selected to distinguish from the other effects.
We classify the sightings as {\em almost certain} (N=5), {\em very probable} (N=4), {\em probable} (N=3),
{\em very possible} (N=2), {\em possible} (N=1) or {\em potential} (N=0),
with N being the number of criteria fulfilled: \\
(i) colour, e.g. reddish, fiery, blood(y), scarlet, green, blue, or the report of a {\em dragon/snake}
or {\em war armies} (both assumed to have some non-white colour), etc.,
while wordings just as {\em white, brilliant,} or {\em glow} are neither sufficient
nor a contradiction (neutral); \\
(ii) aurora-typical motion (changes, pulses, or strong dynamics) including apparent motion indicated
by the words {\em fire, fiery, fight, (war) armies, dragon(s)}; \\
(iii) northern direction (or intrinsically north by mentioning northern celestial regions)
including NE and NW as well as from {\em E to W} or {\em W to E} 
(reports like, e.g., {\em in the east}, are neutral, but excluding purely southern sightings); \\
(iv) night-time observation (darkness), wordings like {\em after sunset} or {\em before sunrise} would not
neccessarily indicate aurorae, but also do not disprove the possibility of an aurora, so that they
are considered neutral (when medieval authors write, e.g., {\em after sunset}, they mean shortly after,
if not even during sunset, but not during the deep dark night);
if the event clearly happened during civil or nautical twilight,
we cannot conclude that it was an aurora;
when the observing time was given as Chinese night-watches,
we converted to local times; and \\
(v) repetition of the event (even if weaker) in at least one of the next three nights. 

With the parameter $N$, we list the number of criteria fulfilled.
Texts like {\em during dusk} or {\em dawn} or {\em white light} are neutral, 
they neither fulfill one of the criteria, nor contradict an aurora. 
Events with N=0 are listed below the likely true aurorae at the end of Sect. 3.2. 
If certain criteria are not fulfilled, this does not automatically mean that the event is not
an aurora, unless a contradiction is clear (e.g. day-time).
Even if some criteria are fulfilled, but the event is obviously something else (not aurora),
we omit it from our catalog;
it is always neccessary to consider the complete story with its context.
For completeness and transparency, the reports interpreted here as non-auroral are listed in Sect. 3.4.
Some dubious events are still in the catalog,
for which there is no clear contradiction against an aurora --
reservations are given. More information about those events may be found later.

Whether any of our five criteria is fulfilled, depends on the words used. 
In occidental records, words like {\em fire} or {\em burning} indicate both red colour and some (apparent) motion,
while {\em flame(s)} do(es) not neccessarily indicate red colour, but (apparent) weak motion, as confirmed by the
fact that such reports often describe some kind of motion together with flames or fire; 
fire is often connected with the colour of blood, i.e. clearly meaning red colour.
In oriental reports, the phrases {\em like fire} and {\em like flame(s)} 
do not mean red colour nor motion;
most oriental reports are written by astronomers, 
they mention explicitly colour and changes.
The word {\em rainbow} or {\em bow} can mean an aurora, 
but could also indicate a solar or lunar halo display (or a night rainbow).
If we do not have any information on the text of a presumable report about aurorae,
but if that event was classified as aurora before by scholars, we still classify it as
{\em potential}, even though N=0 (listed separately at the end of Sect. 3.2).
European observers often do not mention whether the event took place at night nor do they mention the date,
while the Chinese court astronomers often list details of their {\em vapour} (Chinese: {\em qi}) with exact dates or hours, 
colour, form, directions on sky, stellar constellations, changes, dynamics,
and whether it happened during the day or night.

The number N classifies only the {\em likelihood} that a particular event could be an aurora. 
The {\em strength} of the aurora can be determined by considering
its colour, brightness, dynamics, duration, geomagnetic latitude etc. 
Aurorae with N$\ge 3$ tend to be strong.

Further below (Sect. 3.4), we also list all those observations from AD 731 to 825, 
which were previously presented as aurorae in astronomical literature, 
but which we find to be better explained as some different phenomena.

We try to list each aurora event on a certain date only once, 
even if reported in several sources (often depending on each other) with similar wording.
However, if there are different reports with different wording and different information
about the event (in particular from different countries or even continents), 
then we list them separatly, even if they happened on the same day\footnote{The two
entries listed below for AD 743 June from Constantinople and Syria/Arabia
are plotted as two events in Figs. 1 and 2 here, but are plotted as one event in
Neuh\"auser \& Hambaryan (2014).} -- the fact that
independent sources from different countries are available increase the
likelihood of the event to be an aurora in any case, and may also indicate a stronger event.
If there are similar reports about an aurora from, e.g. Ireland 
and continental Europe, for a certain year, but one or both without the exact date, 
we cannot simply assume that they refer to the very same date
(so that they are also listed separately) -- unless the wording is (almost)
identical or it is known that one source depends on the other.

\subsection{Aurora catalog AD 731 to 825}

We now present our aurora catalog from AD 731 to 825.
Reference codes are listed at the end of the listing. We list here only a translation to English (sometimes
our translation), while we will present the European texts in their original language (and English translation) 
and others in (partly improved) English translations in the larger version of
the catalog (AD 550-845) together with more detailed discussion about dating and reliability
later (in prep.). 
For the East Asian events from AD 757 to 776, a detailed discussion about the
datings, the hours in the night, the sizes, the wordings, the
contemperous interpretations, and the translations (together with the
Chinese texts in Chinese) can be found in Chapman et al. (to be submitted).
Below, we give the place (city or area), where the aurora was observed, if known.
If a particular event was listed in one of the large recent aurora catalogs
(Link 1962, Keimatsu 1973, 1974, Yau et al. 1995, Xu et al. 2000),
we mention this explicitly (also in Sect. 3.4).
\begin{itemize}
\item (no aurorae in any catalog found from AD 719 to 733)
\item AD 734, Constantinople (Turkey): {\em On the sky there appeared a fiery shining sign} 
(Theophanes for AD 734, as noticed by us, L62, 
D79, and H96 
dated to AD 735 from Lycosthenes and Frytsch) \\
N=2, i.e. very possible 
\item AD 735, Ireland: {\em A huge dragon was seen, with great thunder after it, at the end of autumn} 
(MAMN83 and MB97 from Ulster Chronicle) \\
N=2, i.e. very possible ({\em dragon} implies colour and apparent motion, 
but could also be some halo (arc) effect with a thunderstorm,
while {\em thunder} could be aurora sound)
\item AD 743 June, Near East (Syria, Arabia): 
{\em A mighty sign appeared in the heavens like columns of fire blazing
in June and stayed and this was the first ...}
(Cook 2001 and S84 from Agapius of Manbij,
similar in Mi\-chel le Syrien in D79, 
dated AD 742 June {\em and} 745 by H96 from D79, dated to AD 742 in Italy by S98) \\
N=2, i.e. very possible (Cook 2001 interpreted this as comet, 
while Cook (1999) argued that this event in June is different from the
event reported for winter 744/5, where a comet was observed in the near and far East.)
\item AD 743 June, Constantinople (Turkey): 
{\em In June, a sign appeared on the northern sky} 
(Theophanes, as noticed by us, most certainly AD 743 following Rochow (1991),
but AD 742 in Breyer (1964), 
F73 and L62 from Cedrenus and Historia miscella) \\
N=1, i.e. possible (maybe the same event as before, but with 
information about northern direction)
\item AD 743 Sep, Near East: 
{\em ... and this was the first [see above for AD 743 June Near East], then another 
appeared in Sep\-tem\-ber like a flame of fire and spread from the East to the West}
(Cook 2001 from Agapius of Manbji, similar in Michel le Syrien, as noticed by us) \\
N=3, i.e. probable (Cook 2001 interpreted this as comet, but see remark to the event
AD 742 June Near East)
\item AD 744, Constantinople (Turkey): {\em This year, a sign appeared on the northern sky} 
(D79 from Theophanes, F73 and L62 from Cedrenus and Historia miscella,
AD 744 in Rochow (1991), but AD 743 in Breyer 1964) \\
N=1, i.e. possible
\item AD 745 England: {\em 744 ... and in the following year, 
fiery strokes (ictus) were beheld in the air, such as no men of that generation had ever seen before, 
and were visible throughout almost all the night of Jan 1} 
(L62 from Roger Hoveden, Symeon Dunelmensis,
and Flowers of History of Matthew of Westminster, 
also F73 from Ara\-go from Perry, and S98 from Britton (1937) and Lowe (1870),
dated to AD 748 Jan 1 in Lowe 1870, 
given for 743 Jan 1 {\em and} 745 Jan 1 in H96, both from D79) \\
N=3, i.e. probable, new moon 745 Jan 7 (Latin {\em ictus} can mean stroke or blizzard or rays,
all consistent with aurora, possibly even pulses)
\item AD 745 Ireland: {\em A horrible and wonderful sign was seen in the stars at night} 
(MAMN83, and MB97 from Ulster Chronicle) \\
N=1, i.e. possible (maybe the same event as in AD 745 England above)
\item AD 745 Syria: {\em And at the beginning of Kanon II (Jan) ... In the middle of the sky, 
it was seen during the night something like a large column of fire} 
(Mi\-chel le Syrien) \\
N=3, i.e. probable (the wording {\em at the beginning} in near eastern reports 
typically mean the 1st of the month; 
Mi\-chel le Syrien used the Seleucid calendar in its west Syiran variation, 
a solar calendar with month like in the Julian calendar; 
{\em Kanon II} is January;
hence, probably simultaneous with the AD 745 event in England)
\item AD 746 Ireland: {\em Dragons were seen in the sky.} (MB97 and MAMN83 from Ulster Chronicle) \\
N=2, i.e. very possible ({\em dragons} imply colour and apparent motion, 
but could also be some halo arc)
\item AD 757 Feb 20, Nan-yang, China: {\em Chih-te reign-period, 2nd year, 
1st month, day ping-tzu (13). At Nan-yang, at night,
there were four white rainbows. They stretched upwards for more than 100 chang [100 chang = 303 m].} 
(K73, YSW, and XPJ from Hsin-T'ang-shu) \\
N=1, i.e. possible, new moon Feb 22/23 (hence, neither halo nor night rainbow) 
\item AD 760 Jul/Aug, China: {\em Ch'ien-yuan reign-period, 3rd year, 6th month (XPJ: Jul 17 - Aug 15). 
At dusk, there were 3 blue vapours in the NW.}
(DC80, YSW, and XPJ from Hsin-t'ang-shu) \\
N=2, i.e. very possible (but maybe a halo effect at sunset (NW), {\em at dusk} is neutral, 
i.e. neither clearly indicating night nor excluding darkness, but NW direction is typical
for likely true Chinese aurora reports)
\item AD 762 May 1, Xian, Shaanxi, China: {\em Pao-ying reign-period, 1st year,\footnote{given
as {\em second year} in YSW: It was the 2nd year since the end of the
reign period of the previous Emporer Shangyuan,
but the first year of the reign period of the new Emporer Pao-ying 
(J. Chapman, priv. comm.).} 4th month, day jen-tzu (49). 
At night, a red light like flames was seen in the NW. Its blazing flames streched across the sky 
and penetrated Tzu-wei [circumpolar].
It gradually floated towards E and spread to the N. It shone brilliantly for several tens of li. 
After a long time then it was dispersed.}
(K73, YSW, and XPJ from Chiu-t'ang-shu, T'ang-hui-yao) \\
{\em Several tens of li} are several times 5-6 km.
There is another report about an aurora with almost the same wording
(but leaving out the exact date and the night-time) at a slightly later time in the same year
(AD 762 between Jul 26 and Aug 23)
from Wenxian tongkao (J. Chapman, priv. comm.)
listed in DC80 and in S98 for AD 762 Aug for Xian, China. 
It may well be a misplaced duplication, so that we do not list it separately. \\
N=4, i.e. very probable, new moon Apr 28/29 
\item AD 762 May 20, Chiang-ling-hsien, Hu-pei, China: \\ 
{\em Pao-ying reign-period, 1st year,\footnote{see previous footnote} 
4th month, day hsin-wei (8). At night, at Chiang-ling
(in Hu-pei Province) a red light was seen. It penetrated Pei-tou [Great Dipper near $\alpha$ UMa].} 
(K73, YSW, and XPJ from Chiu-t'ang-shu) \\
N=3, i.e. probable, new moon May 27/28 (Great Dipper is north)
\item AD 762, Ireland: {\em A bright night in autumn.} (MAMN83, N72, 
and H96 from Ulster 
and Scotish Chronicle) \\
N=1, i.e. possible (dating uncertain, as also listed for AD 710 in the Scotish Chronicle, 
and AD 713 in Hennessy (1964), N72: {\em AD 710 or 714 in England})
\item AD 762 Sep 16, Xian, Shaanxi, China: {\em Pao-ying reign-period, 
1st year,\footnote{see previous footnote} 8th month, day keng-wu (7). 
At night, there was a red light stretching
and penetrating Txu-wei [circumpolar]. It gradually moved towards the NE and 
permeated half of the sky.}
(K73, YSW, and XPJ from Hsin-t'ang-shu, Chiu-t'ang-shu) \\
N=4, i.e. very probable, new moon Sep 22 (maybe (qua\-si-)simultaneous with the previous event;
Jiu Tang shu no. 11.270 has {\em red light in the NW} and then moving {\em gradually towards the NE}
(J. Chapman, priv. comm.), while YSW and XPJ mention only the NE)
\item AD 764, Ireland: {\em A bloody flux throughout Ireland. \\
Three showers fell 
in Crich Muiredaig in Inis Eogain, i.e. a shower of pure silver, a shower of wheat,
and a shower of honey. The three showers of Ard Uilinne. Fell from heaven for love of Niall: 
A show\-er of silver, a show\-er of wheat.
And a show\-er of honey. Fergal's manly son was dubbed for this among warriors: 
Since everyone came to follow them Niall of the Showers is his name.}
(MAMN83 from Ulster Chronicle) \\
N=2, i.e. very possible (listed with similar wording also for AD 718 in Ulster Chronicle;
while the {\em bloody flux throughout Ireland} could be an aurora 
(or reddish pollen or dust, e.g. from the Sahara desert),
other early suggestions include red algae (Rhodophyta) in rivers, 
reddish liquid drops from many butterflies leaving their cocoons, 
or blood drops from many bees leaving their beehive (see e.g. Schwegler 2002),
one of the {\em three showers} could be a natural real wheat shower -- wheat first blown up
by wind, then falling down elsewhere)
\item AD 765, England: {\em In the year 765, fiery strokes were seen in the air, 
such as formerly appeared on the night
of Jan 1, as we have already mentioned [AD 745 Jan 1].} 
(L62 from Master of Rolls from Roger de Hoveden, 
N72 and H96 from Melrose, 
also listed in U13) \\
N=3, i.e. probable
\item AD 765, Ireland: {\em A horrible and wonderful sign was seen in the stars at night.} 
(U13 and MAMN83 from Ulster Chronicle) \\
N=1, i.e. possible (maybe same event as AD 765 above;
entry with same wording in both AD 745 and 765 in MAMN83;
MB97 list them for AD 744 and 764 and accept the entry for AD 744 as more likely)
\item AD 765, Europe: {\em Fires were seen on sky.} (P88 and F73 from Toaldo, Vogel, Ara\-go, Perrey) \\
N=2, i.e. very possible (maybe same event as the previous two items for AD 765)
\item AD 767 Oct 8, Xian, Shaanxi, China: {\em At night white mist (wu) arose in the NW of the wei (in Sco). 
It permeated and spread over the sky.} 
(K73 from Chiu-t'ang-shu, also listed in YSW and U13) \\
N=2, i.e. very possible, full moon Oct 12 (reported direction and stellar constellation speak 
for aurora, while mist can also indicate fog)
\item AD 770 Jun 20, Xian, Shaanxi, China: {\em Ta-li reign-peri\-od, 5th year, 5th month, day chia-shen (21). 
In the NW, a white vapour extended across the sky.}
(K73 and YSW from Chiu-t'ang-shu, also listed in U13) \\
N=1, i.e. possible, moon's last quarter Jun 20/21 (see comments on the next event)
\item AD 770, Jul 20, Xian, Shaanxi, China: {\em Ta-li reign-peri\-od, 5th year, 6th month, day chia-yin (51). 
A white vap\-our appeared in the NW direction. It extended accross the sky.}
(K73 and YSW from Chiu-t'ang-shu, also listed in U13) \\
N=1, i.e. possible, moon's last quarter Jul 20 (the Chinese characters for the dates 
given on the sexagenary cycle for this event (chia-yin for day 51) 
and the previous event (chia-shen for day 21) could be mixed up, so that one of the
two events could be a misdated duplication of the other; however, since the texts are
also slightly different, we assume that these are two different (similar) events;
even though night-time is not mentioned, this event is about one solar rotation period 
after the previous one, i.e. possibly recurrent activity due to a stable coronal hole;
these two events are reported for the NW, the typical direction of Chinese aurora sightings)
\item AD 772, Amida (now Diyarbak\i r, Turkey): {\em Another sign appeared in the northern side,
and its appearance gave testimony about the thread and menace of God against us.
It was seen at harvest time, while occupying the entire northern side, from the east end to the
west end. Its likeness was as follows: a red sceptre, a green one, a black one, 
and a yellow one. It was moving
up from the ground, while one sceptre was vanishing and another appearing. When someone looked at it,
it would change into 70 shapes. For the intelligent person the sign indicated menace. 
Many people said many things about it; 
some said it announced bloodshed, and others said other things. But who knows the deeds
of the Lord ? -- I will give signs in the heaven, and wonders on the earth. [from Bible Acts 2, 19]} 
(Harrak 1999 from the Chronicle of Zuqnin) \\
N=3, i.e. probable, timing in the year AD 771 Oct to 772 Sep at harvest 
(Chabot 1927 gave {\em la moisson} (corn harvest) in his French translation 
of the Chronicle of Zuqnin; A. Harrak confirmed (priv. comm.)
that the Syriac word for {\em harvest} means ({\em corn harvest})
if not otherwise specified), i.e. summer AD 772; 
S84 and D79 misdated to AD 766; 
the additional event of AD 773 is correctly dated (see below, {\em sign that was seen a year ago}), 
so that this event must be AD 772; Neuh\"auser et al. (in prep.) will give more details 
about this chronicle and its celestial observations with many original drawings.
\item AD 772 (around Sep 29), Ireland: {\em The assembly of the hand-clapping 
at which occurred lightning and thunder like the day of judgment.
The hand-clapping on St Michael's Day 29 Sep which called fire from heaven.} 
(U13 and MAMN83 from Ulster Chronicle) \\
N=2, i.e. very possible, new moon Oct 1, but maybe just thunderstrom with blizzard(s)
\item AD 773, Amida (now Diyarbak\i r, Turkey): 
{\em The sign that was seen a year ago in the northern region was seen again in this year, 
in the month of Haziran (June), on a Friday. 
It was on Fridays that it used to appear during these three consecutive years,
stretching itself out from the eastern side to the western side. 
When seen by someone, it would change into many shapes,
in such a way that as soon as a red ray vanished, a green one would appear, 
and as soon as the green one vanished,
a yellow would appear, and as soon as this one vanished, a black one would appear.} 
(Harrak 1999 from the Chronicle of Zuqnin) \\
N=3, i.e. probable, timing in the year AD 772 Oct to 773 Sep at Haziran, i.e. AD 773 June
(the chronicler of Zuqnin probably used the west Syrian Christian version of the Seleucid calendar as it 
was in use in that area since the 7th century, 
where the year runs from Oct 1 to Sep 30, and the months are defined as in the
western (Roman) solar calendar, not as in the lunar or luni-solar calendar; 
hence, the month called {\em Haziran} would exactly be our June);
S84 misdated to AD 770, D79 misdated to AD 770-772 (see also Sect. 3.3);
the report gives one event per year for three consecutive years including this year 
clearly datable to 773 (and the previous report AD 772 given the start of the text for AD 773:
{\em The sign that was seen a year ago ...});
a 3rd text is missing, possibly due to incompleteness of the manuscript.
\item AD 786, Europe: {\em and blood flow from sky and earth} (S64 from An. Chesnii,
also in An. Lamberti and Weissemburg) \\
N=2, i.e. very possible (maybe the same as the next; but see also the
alternative explanations for {\em bloddy flux} for AD 764 Ireland)
\item AD 786, Europe: {\em In this year in the month of December, terrible war armies appeared 
in the sky as they never appeared previouly, in our times.} (L62 from Moissiacense,
F73 from Quetelet, Sch\"oning (1760), S64 from Lauresheim, 
the wording {\em as they never appeared before} may indicate a longer period
without strong aurorae before AD 786) \\
N=2, i.e. very possible ({\em war armies} can mean red colour and motion)
\item AD 786 Dec 21, Xian, Shaanxi, China: {\em Emporer Dezong of Tang, 2nd year, 
of the Zhenyuan reign period, 11th month, day renzi (49).
After sunset there were five scarlet vapours that emerged from black clouds and spread over the sky.} 
(K73 and XPJ from Xin Tang shu, Wuxing zhi) \\
N=1, i.e. possible, new moon Dec 25; as reported in Xu et al. (2000),
instead of {\em day renwu (19)} as in K73, it is {\em day renzi (49)}, hence Dec 21;
the wording {\em after sunset} is neutral and may indicate twilight
(maybe quasi-simultaneous with the previous event)
\item AD 793 (before Jun 8), Northumbria, England: {\em In this year terrible portents appeared
over Northumbria, and miserably frightened the inhabitants:
these were exceptional high winds and flashes of lightning, and fiery dra\-gons were seen flying in the
air. A great famine soon followed these signs; and a little after that in the same year on 8 Jan, the
harrying of the heathen miserably destroyed God's church in Lindisfarne by rapine and slaughter.}
(J70, S84, N72, L62 from Anglo-Saxon Chro\-nicle, also in Sch\"oning 1760
and H96)
N=2, i.e. very possible (correct date for the attack of the Vikings is AD 793 Jun 8
according to Simeon of Durham,
so that it may have been (quasi-)simultaneous with the events 
listed next (Arabia, May);
the plural {\em portents} above may indicate more than one sighting,
{\em flashes of lightning} could be pulses)
\item AD 793 May 11/12 \& 14/15, Iraq: {\em In this year} (177 Hijra, i.e. 793 Apr 18 - 794 Apr 6), 
{\em according to what al-W\={a}qid\textit{\={\i}} has mentioned, there
occurred a (violent) wind} [ar.: {\em \d{h}\textit{\={\i}}r}], 
{\em and overshadowing (of the heavens)} [ar.: {\em \d{z}ulma}] {\em and a redness (in the sky)} 
[ar.: {\em \d{h}umra}], 
{\em on the night of Sunday} (i.e. the night of Satur\-day-Sunday), 
{\em the 26th of al-Mu\d{h}arram} (793 May 13). 
{\em Then there was a further overshadowing (of the heavens) on the night of Wednesday} 
(i.e. the night of Tues\-day-Wednesday), {\em the 28th of al-Mu\d{h}arram} (793 May 15), ...
(Bosworth 1989 from al-\d{T}abar\textit{\={\i}}, date conversions given in brackets 
from Bosworth (1989).
However, given the intervals between the weekdays and the Arabic dates given (here and
in the next event), the Julian dates from Bosworth (1989) are intrinsically strongly inconsistent,
while al-\d{T}abar\textit{\={\i}} himself may be only slightly inconsistent.
Assuming that the Islamic lunar month has either 29 or 30 days, 
we try to find a solution with the least number of modifications, namely just one:
The first sighting on Sunday night (the night of Satur\-day to Sunday) dated 26 Mu\d{h}arram was on May 11/12;
the sighting reported for the night from Tuesday to Wednesday was on May 14/15, which was
indeed a night from Tuesday to Wednesday (Spuler \& Mayr 1961), but 29 Mu\d{h}arram instead of 28; 
the last sighting in the night from Thursday to Friday (next event) reported for 2 \d{S}afar 
was in the night of May 16/17, so that the month of Mu\d{h}arram had 29 days.
The month of Mu\d{h}arram then started already on AD 793 Apr 16/17 (evening of Apr 16
with the first sighting of the crescent new moon), which is well possible given
the true conjunction (new moon) on AD 793 Apr 15 at 6:25h UT
(in the calculated Islamic calendar for the year 177 Hijra, the month of Mu\d{h}arram would have started on AD 793 Apr 17/18
and the next month of \d{S}afar on AD 793 May 17/18, but the calculated Islamic calendar has an uncertainty of $\pm 2$ days,
partly due to the unknown date of the true first lunar crescent sighting in each month, 
see, e.g., Spuler \& Mayr 1961 and Neuh\"auser \& Kunitzsch 2014). \\
N=4, i.e. very probable, new moon May 14 at 18h UT
\item AD 793 May 16/17, Iraq: {\em ... and then a violent wind and intense 
overshadowing of the heavens on Friday, the 2nd of \d{S}afar} (793 May 19).
(Bosworth 1989 from al-\d{T}abar\textit{\={\i}}, date conversion given in brackets from Bosworth (1989);
however, given the discussion for the previous event, the last event was on 16/17 May, 
the night of Thursday to Friday) \\
N=2, i.e. very possible, new moon May 14 at 18h UT
\item AD 796 Oct 20, Xian, Shaanxi, China: {\em Cheng-yuan rei\-gn-period, 12th year, 9th month, day kuei-mao (40). 
At night, there was a red vapour like fire. 
It was seen in the N direction with its upper part 
reaching Pei-tou [Great Dipper near $\alpha$ UMa].} 
(K73, YSM, and XPJ from Hsin-t'ang-shu and T'ang-shu) \\
N=3, i.e. probable, full moon Oct 20/21 (this example shows that 
an aurora in the north can be seen even during a full moon night)
\item AD 800, Europe: {\em red sign with many fiery lances on sky} (F73 from Frobesius from Frytsch) \\
N=2, i.e. very possible (but also somewhat problematic, 
because neither L62 nor we found the original source,
so that the dating is more uncertain).
\item AD 804 Oct 25, Chang-an, China: {\em 19-21h o'clock, there was a white vapour touching the sky 
at the east and the west} (K74 from T'ang-shu) \\
N=2, i.e. very possible (full moon Oct 22, so that there was no moon {\em at the east} in the evening of Oct 25;
sunset at Chang-an (the modern Xian) on that date was at 16:54h local time, 
end of astronomical twilight at 18:19h local time, both before the 19-21h double-hour)
\item AD 807 Feb 26, Germany: {\em Again on the 4th calends March was an eclipse of the moon, 
and war armies appeared that night of an amazing brightness} 
(M33, P88, S64, 
N72, and H96
from Royal Frankish Annals, Chronicle Loisel and many others, 
English translation from N72) \\
N=3, i.e. probable (full moon Feb 25/26 with total lunar eclipse, sunspot then AD 807 Mar 17-24)
\item AD 817 Oct, Germany: {\em in same year fiery war armies appeared in sky in October} 
(L62,
N72, and H96
from Xanten Chronicle) \\
N=2, i.e very possible
\item AD 817 Oct 29, Iraq: {\em On Friday night, a reddish glow appeared 
in the sky and stayed until late at night.
Later on it disappeared but two red columns remained until the dawn} 
(Rada \& al-Najeh 1997, Basurah 2006, both from Ibn Al-Ath\textit{\={\i}}r, the former also give the Arabic text);
the very similar, but reduced report saying
{\em a reddish glow appeared in the sky and stayed until late at night}
from a chronicle from Yemen from Ibn al-Qasim (see Basurah 2005)
is probably derived from the Iraqi report on the same event
(Ibn al-Ath\textit{\={\i}}r is much closer in time), even though it is reported for the
Islamic lunar calendar year from AD 816 Aug to 817 July, i.e. before the date of the Iraqi event
(also not listed for Yemen in Table 1) \\
N=3, i.e. probable (full moon Oct 28, (quasi-)simultane\-ous with the previous event in Germany)
\item AD 819 Jan 6, Xian, Shaanxi, China: 
{\em There was a white rainbow, 50 degrees wide, extending from east to west over the sky}
(K74 from T'ang-shu) \\
N=1, i.e. possible (moon's first quarter Jan 6/7)
\item (no aurorae in any catalogs found from AD 820 to 825)
\end{itemize}
References are M33 (Mairan 1733), P88 (Pilgram 1788), F73 (Fritz 1873),
L62 (Link 1962), S64 (Schove 1964), N72 (Newton 1972),
K73 (Keimatsu 1973), K74 (Keimatsu 19\-74), D79 (Dall'Olmo 1979), DC80 (Dai \& Chen 1980), 
MAMN83 (Mac Airt \& Mac Niocaill 1983),
S84 (Schove 1984), YSW (Yau et al. 1995), MB97 (McCarthy \& Breen 1997), 
H96 (Hetherington 1996),
S98 (Silverman 1998 online aurora catalog),
XPJ (Xu et al. 2000), U13 (Usoskin et al. 2013).
The recent work by Chai \& Zou (2014) does not present any new aurora observations.

We now also list the four events with N=0, i.e. where none of our
aurora criteria are fulfilled nor were any contradictions found, according to the texts known,
but which were classified as aurora by other scholars before,
so that we consider them {\em potential} 
(i.e. not considered to be likely true aurorae):
\begin{itemize}
\item AD 740 Germany: {\em Signs in Sun and moon and in stars appeared} 
(L62 and F73 from Xanten, Remense, Fredegar, and Mettuenses,
dated AD 741 Feb or Mar in F73, but AD 740 in L62, 
740/741 in D79 and H96) \\
N=0, i.e. potential (could refer to Lk 21, 25 in the New Testament)
\item AD 786, Ireland: {\em A horrible vision in Cluain Moccu Nois, and great penance done 
throughout Ireland} (S64, MAMN83, and U13 from Ulster Chronicle) \\
N=0, i.e. potential
\item AD 788 Sep 23 (approximate), Northumbria, England: {\em In this year Aelfwald, king of Northumbria, 
was slain by Sicga on 23 Sep, and a light was frequently
seen in the sky where he was slain: he was buried at Hexham inside the church.} (Anglo-Saxon Chronicle) \\
N=0, i.e. potential, not neccessarily several nights, may\-be several times in one night, full moon Sep 20
(dating uncertain, on or after Sep 23 in AD 788 or 789)
\item AD 819, England: {\em heaven afterwards revealed the deed by means of a column of light}
(J70 from Florence of Worcester),
interpreted as aurora by J70, but possibly only a halo pillar \\
N=0, i.e. potential
\end{itemize}

Given that we have 39 likely true aurorae from AD 731-825,
but only one sunspot, aurorae are better suited here for the
reconstruction of solar activity than spots.
In Fig. 1, we present the distribution of solar activity proxies for the time from AD 731 to 844 
(i.e. around AD 775) including the aurorae listed above. 
Aurorae with N=0 are not plotted in Figs. 1 \& 2, 
they were also not plotted in Neuh\"auser \& Hambaryan (2014).
All aurorae from AD 550 to 844 will be presented
in a later publication, together with most original texts.

\subsection{Discussion: Strong aurorae around AD 775?}

We will discuss now whether there are any relatively strong aurorae 
around AD 775, also by comparison to the decades before and afterwards --
strong aurorae can be recognized by low geomagnetic latitude,
simultaneous observations in different countries, repetition in several nights,
strong dynamics and/or colour.

We notice several aurorae sightings in Amida at the Tur\-kish-Syrian border in the early 770s.
A problem with the reports from the {\em Chronicle of Zuqnin} 
(old Syriac with modern French translation 
in Chabot 1927 and English in Harrak 1999) is the dating: 
This chronicle runs until its ends in AD 775/6, 
but significant parts of extended reports about the last few years are 
worm-eaten (Chabot 1927, Harrak 1999);
aurorae are mentioned for three consecutive years not before AD 770, 
but the detailed reports are found for only two of such events, 
which were most certainly AD 772 and 773, both in the summer.
The latter is clearly datable to summer AD 773 and the former was {\em a year ago}.
Previously, other scholars gave other dates, 
probably partly because the author in a few cases goes back and forth in time.
These aurorae were not mentioned in U13. 
One could consider whether these are strong aurorae related to a solar super-flare.
However, the author of the {\em Chronicle of Zuqnin} is strongly apocalyptic, 
who interprets celestial observations -- following the Bible -- as portents,
so that the descriptions are always intermixed with interpretation.
In case of these aurorae, he describes both as dynamic and dramatic,
his drawings of these aurorae show that they are mainly colourful aurora curtains
(the drawings will be published in Neuh\"auser et al., in prep.).
Some of the descriptions of the 2nd sighting (AD 773) seem to be
repetitions of the report for AD 772.

For AD 772, we have another report from Ireland saying: \\
{\em The assembly of the hand-clapping at which occurred lightning and thunder like the day of judgment.
The hand-clapp\-ing on St Michael's Day 29 Sep which called fire from heav\-en}, from the Ulster Chronicle,
which is more difficult to interprete, as it could also be thunderstorm with blizzard(s),
but was also interpreted as aurora in U13. We classify it as {\em very possible aurora} with $N=2$,
new moon was AD 772 Oct 1, close to the date of St Michael's Day and this event.
While the Chronicle of Zuqnin says that the aurorae happened on a Friday during (corn) harvest AD 772, 
the day given in the Ulster Chronicle (AD 772 Sep 29) was a Wednesday.
Since (corn) harvest in Syria/Turkey takes place much earlier than Sep 29, 
the events were not simultaneous.

For the period studied here, we see strong aurorae in the following years: \\
(Quasi-)simultaneous reports about aurorae from different countries 
were found for AD 745 (England, Syria), AD 762 (Ireland, China),
AD 786 (Europe, China), AD 793 (Iraq, England),
and AD 817 (Germany, Iraq). The aurora from AD 807 Feb 26 is
about half a month before the sunspot sighting AD 807 Mar 17-24,
both in the Royal Frankish Annals. \\
There is one aurora report with repetition of the sighting within
a few days,
presumably due to a CME, which was in AD 793 (Iraq). \\
Stronger dynamics were noticed in aurorae in
AD 762 May 1 (China), AD 772/3 (Amida, see above), and AD 793 (Iraq, England). \\
There were three aurorae with color other than red,
name\-ly in AD 760 ({\em blue vapour}) in China
(but maybe a halo effect) and then red-yellow-green aurorae 
in AD 772 and 773 in Amida, now at the Turkish-Syrian border.

\begin{table*}
\begin{center}
\caption{{\bf Geo-magnetic latitude of low-latitude aurorae AD 731-825.}
We list first the geomagnetically lowest-latitude aurorae from AD 731 to 769,
then all aurorae found from AD 770 to 780,
and then the geomagnetically lowest-latitude aurorae from AD 781 to 825.
We give the year in which the aurora was observed,
the area (country as of today) where it was observed,
the reference where the report was found (abbreviations as given at the end
of Sect. 3.2 and B06 for Basurah (2006) and Rada \& al-Najeh (1997), NN for this work),
the number of aurora criteria fulfilled by that report (Sect. 3.2),
the geographic latitude $b$ (north) for the exact location
given in the aurorae report (or the capital of that country at that time)
and longitude $l$ (west of Greenwich, unless otherwise specified), 
and then the geomagnetic latitude of those aurorae -- 
considering five reconstructions for the location of the geomagnetic pole
near the northern geographic pole: 
Arch3k: Donadini et al. 2009, 
Cals3k3: Korte et al. 2009, 
Cals3k4: Korte \& Constable 2011, 
Cals10k1b: Korte et al. 2011,
Pfm9k1a: Nilsson et al. 2014.}
\begin{tabular}{lll|c|cc|ccccc} \hline
Year & Area & Ref & N & \multicolumn{2}{c}{geographic} & \multicolumn{5}{c}{geomagnetic latitude in degree} \\
AD   &      &     &   & $b [^{\circ}]$ & $l [^{\circ}]$ & Arch3k & Cals3k3 & Cals3k4	& Cals10k1b & Pfm9k1a \\ \hline
\multicolumn{11}{c}{Geo-magnetically lowest aurorae AD 731-769} \\ \hline
757  & China & K73  & 1 & 33.1&112.5&34.3&36.8&37.5&36.1&37.6 \\
760  & China & YSW  & 2 & 34.3&108.9&36.1&38.2&38.9&37.6&39.5 \\
762  & China & K73  & 4 & 34.3&108.9&36.2&38.2&38.9&37.6&39.5 \\
762  & China & K73  & 3 & 34.3&108.9&36.2&38.2&38.9&37.6&39.5 \\
762  & China & K73  & 4 & 30.3&112.2&31.9&34.1&34.8&33.4&35.1 \\
767  & China & K73  & 2 & 34.3&108.9&36.4&38.1&38.9&37.7&39.7 \\ \hline
\multicolumn{11}{c}{All aurorae AD 770-779} \\ \hline
770  & China & K73  & 1 & 34.3&108.9&36.6&38.1&38.9&37.7&39.8 \\
770  & China & K73  & 1 & 34.3&108.9&36.6&38.1&38.9&37.7&39.8 \\
772  & Turkey & D79  & 3 & 41.0&29   &48.9&43.1&43.5&44.5&50.1 \\
772  & Ireland & U13 & 2 & 53.3&6.3 E&58.5&52.9&52.8&54.3&57.3 \\
773  & Turkey & D79  & 3 & 41.0&29   &48.9&43.1&43.5&44.5&50.1 \\ \hline
\multicolumn{11}{c}{Geo-magnetically lowest aurorae AD 780-825} \\ \hline
786  & China & K73  & 1 & 34.3&108.9&37.2&38  &38.7&37.9&40.2 \\
793  & Iraq & NN    & 4 & 33.3&44.5 &41.8&36.3&36.8&37.6&43.9  \\
793  & Iraq & NN    & 2 & 33.3&44.5 &41.8&36.3&36.8&37.6&43.9  \\
796  & China & K73  & 3 & 34.3&108.9&37.7&37.9&38.7&37.9&40.4 \\
804  & China & K74  & 2 & 34.3&108.9&38.1&37.9&38.7&38  &40.6 \\
817  & Iraq & B06  & 3 & 33.3&44.5 &42.7&36.3&36.8&37.6&44.3 \\
819  & China & K74  & 1 & 34.3&108.9&38.9&37.9&38.7&38.0&40.8 \\ \hline
\end{tabular}
\end{center}
\end{table*}

Geo-magnetically relatively low-latitude aurorae show strong solar wind.
There were (so-called) low- to mid-lati\-tu\-de aurorae
(below about $43^{\circ}$ geomagnetical latitude)
from AD 731 to 825 (see Table 1):
We see aurorae in AD 793 and 817 (both Iraq) at a
geomagnetical latitude of $36$ to $44^{\circ}$ 
(considering five different reconstructions of the geomagnetic pole, namely from
Donadini et al. 2009, Korte et al. 2009, Korte \& Constable 2011, Korte et al. 2011, 
and Nilsson et al. 2014);\footnote{To be conservative,
we do not consider the aurora reported for Yemen for AD 816/817 here,
even though it would be at a geomagnetical latitude of $19$ to $27^{\circ}$ only,
because this report could be just derived from the one for Iraq,
so that the observation was probably not in Yemen;
also not listed in Table 1.}
it is not a surprise to find aurorae at the longitude range of Iraq
(as well as Turkey, Syria, and possibly Yemen) in the time AD 731 to 825,
because the geomagnetic pole was located 
at a geographic longitude of roughly $30$ to $45^{\circ}$ east
from about AD 600 to 900 (according to the re-constructions considered).
There are even lower-latitude aurorae, namely 
in AD 757, 760, 762 (three), 767, 770 (two), 786, 796, 804, and 819,
in China at geomagnetical latitude of $32$ to $41^{\circ}$
(again considering the five different reconstructions).

None of those strong aurorae are connected with additional $^{14}$C peaks --
whether they were strong due to simultaneous sightings,
repetition, strong dynamics, and/or relatively low geomagnetic latitude.
If the AD 775 $^{14}$C variation would have been caused by one or more solar
super-flares, even stronger aurorae should have been reported
(for the same time in different countries).
The lack of reports about super-strong aurorae is most certainly 
not due to missing historic documents, 
because there are continuously reports
about celestial events in the decades before, during, and after the AD 770s
(see Sect. 2.3, 3.2, 3.4).
Even for the period of AD 774-785, where there are no aurora reports at all,
there are other reports, both oriental and occidental.  

There are indeed strong aurorae reported for the time from AD 760 to 773.
However, the mid-latitude aurorae (between about $43^{\circ}$ and about $50^{\circ}$)
from AD 772/3 in Amida (Turkey/Syria) cannot be seen as exceptionally strong
aurorae (no repetition and no (quasi-)simultaneous reports) --
they cannot support a hypothetical solar super-flare.

\begin{figure*}
\begin{center}
{\includegraphics[angle=270,width=15cm]{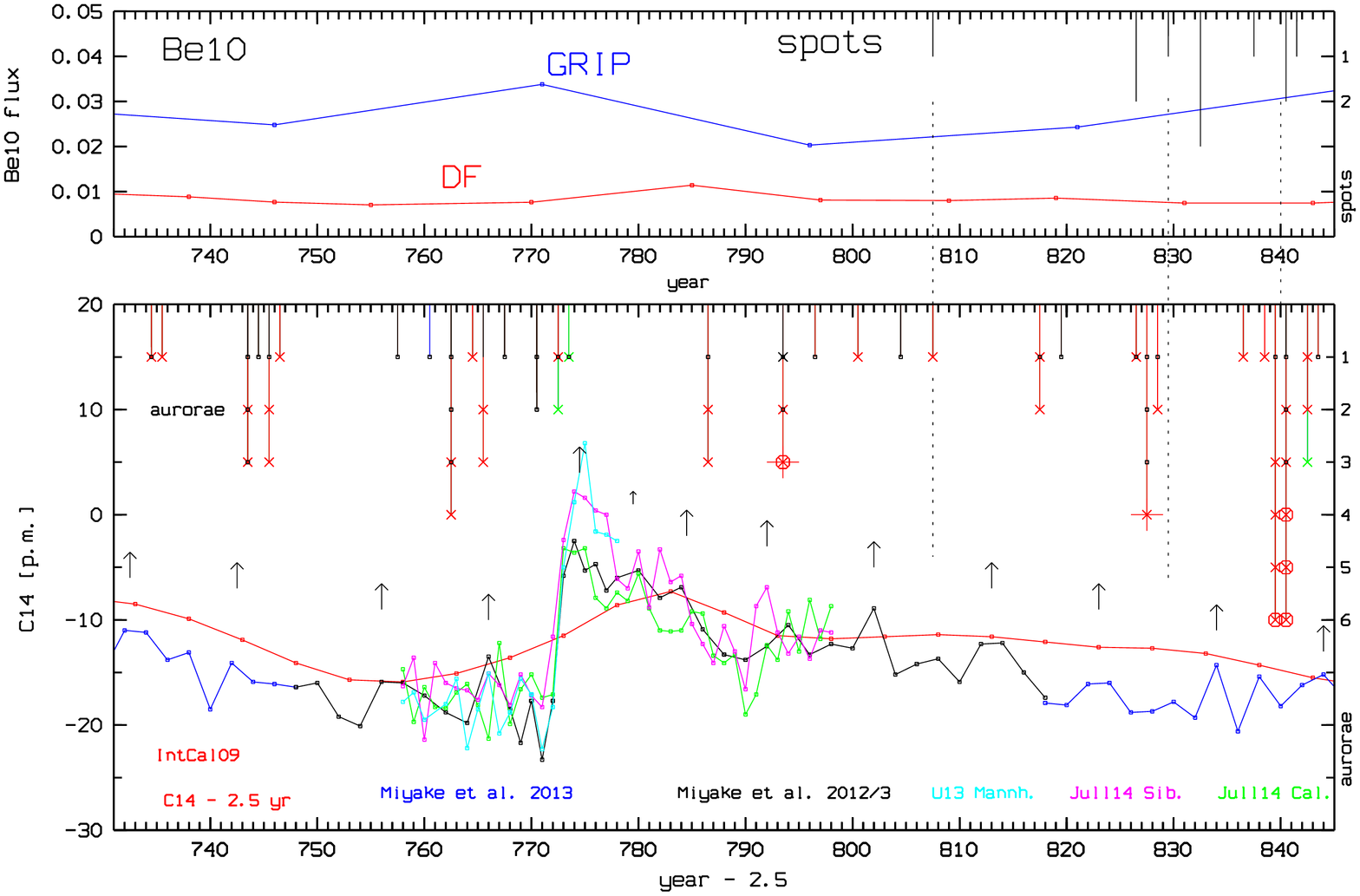}}
\caption{{\bf Solar activity proxies for the time AD 731 to 844:}
Top: $^{10}$Be flux from Dome Fuji (DF) in red (Horiuchi et al. 2008)
and GRIP in blue (Vonmoos et al. 2006, recieved from J. Beer, priv. comm.)
in atoms cm$^{-2}$ s$^{-1}$; 
note that the DF data are matched in time to the radiocarbon data
already in Horiuchi et al. (2008), while the GRIP data are not shifted.
The number of reports about naked-eye sunspots
(see Sect. 2) are indicated as black lines from the top with a scale on the right-hand side y-axis.
Bottom: $^{14}$C (in p.m.) with 5-yr time resolution from Intcal09 (Reimer et al. 2009) in red,
data with better time resolution from Miyake et al. (2012, 2013ab) in black (around AD 775) and dark blue, respectively;
data from Jull et al. (2014) for the Californian and Siberian trees in pink and green, respectively,
data from the Mannheim lab from U13 for a German tree 
in light blue.
Aurorae are shown from our own catalog (Sect. 3.2, 
scale for number of aurorae 
on the right-hand side y-axis, 
potential aurorae with N=0 are not plotted):
Blue, green, and red lines for aurorae with that colour, respectively, black lines for aurorae
where no colour was reported, the length of the lines indicate the number of aurorae for each colour,
correctly coloured crosses (at the end of aurora lines) indicate aurorae where dynamics were reported,
circles indicate aurorae
that lasted for several days (at least two days within four consecutive days, so that the Arabic
aurorae 
in three nights AD 793 May 11/12 to 16/17 are plotted twice, 
once with repetition (May 11/12 \& 14/15) possibly indicating a CME
(and once for May 16/17),
the number of large plus signs indicate the number of aurorae with pulses,
and the number of small black dots indicate the number of aurorae outside
Europe, namely in Byzantium, Arabia, and Eastern Asia,
indicating more southern events.
$^{14}$C data are plotted 2.5 years {\em before} the measured time,
but spots, aurorae, and $^{10}$Be are not shifted.
Spots and aurorae are plotted into the middle of the respective year $x$, i.e. at $x.5$.
For clarity, we omit the $^{14}$C and $^{10}$Be error bars; since the errors are quite similar 
for all data points of a certain data set 
within this time period, there is no significant difference between considering
error-weighted or un-weighted data points for, e.g., dating Schwabe cycles.
We can identify Schwabe cycles by upwards pointing arrows plotted above local $^{14}$C maxima, i.e. activity minima
(the small arrow at the end of the AD 770s indicates 
that there could be two short cycles instead of one).
Dotted lines connect (clusters of) sunspot sightings 
with phases of maximal activity ($^{14}$C minima).}
\end{center}
\end{figure*}

\subsection{Reports misinterpreted as aurora AD 731 to 825}

Given that other celestial phenomema can fulfill aurora criteria,
we have to exclude those presumable aurora reports for which 
a better alternative explanation can be found, e.g. comets or halos;
in addition, we list events where some property observed contradicts our aurora criteria.
For more details about events from East Asia, 
which were misinterpreted as aurorae before,
see Chapman et al. (to be submitted),
which also lists a few more reports, 
which could well be misinterpreted as aurorae.

Next, we list reports, which were misdated and/or misinterpreted as aurora
(with the same abbreviations for the references as listed near the end of Sect. 3.2):
\begin{itemize}
\item AD 743/4, England: {\em This year a red crucifix appeared in the heavens 
after sunset} from J70, 
who took it from Florence Worcester and the Anglo-Saxon Chronicle 
(under AD 743 as stated by J70),
but misdated, correct year is AD 776, see below.
\item AD 752, Italy: {\em It appeared to him [pope Stephanus], while traveling, a large
sign in the sky in one of the nights, namely a fiery ball [globus] in southern direction,
which turned away from the French [Gallia] areas towards the areas of the Langobards [Italy]}
translated by us from Latin in Siegebert Gembl. 
P88 interpreted this as aurora -- partly based on an incorrect German
translation of the Latin text from Siegebert Gembl.
This was probably a bolide (southern direction contradicts aurora).
\item AD 761 Dec 13, Xian, Shaanxi, China:
{\em In that year [the second year of the Dayuan reign period, AD 761], 
on the guiyi day of the 11th month, 
at the hai [double-]hour [21-23h] 
after the first night drum [in] the second fifth-hour [i.e. in the 2nd of five parts
with equal length within the night watch]  
the moon covered Mao [the Pleiades], and then emerged to its north.  
It was surrounded by a white halo. The stars of Bi [in Tau] had white qi [vapour] 
amongst them which followed the moon north to penetrate Mao}
a translation by J. Chapman (priv. comm.) from Jiu Tang shu for Chang’an
(now Xian) in China;
the translation given in K73 is
not that clear (a halo around stars in Tau?), he classified it as {\em probable} aurora,
but his translation is also fully consistent with a halo display. 
This event is also listed as aurora in S98, but neither in YSW nor in XPJ.
The above translation from J. Chapman is self-consistent:
The event reported took place {\em at the hai [double-]hour}, i.e. 21-23h,
{\em after the first night drum}, i.e. in the 2nd night-watch,
and in the {\em 2nd fifth-hour}, i.e.
in the 2nd of five equal subsections of the 2nd night-watch;
at Xian, China, on that date, the 2nd night-watch ran
from 20:03h to 22:38h local time, 
so that the 2nd {\em fifth-hour} ran from 20:34 to 21:05h;
hence, the event took place (or started) around 21h.
The moon indeed occulted the Pleiades (Mao) at that time in that night,
later (in the middle of the night) the moon was located north of the Pleiades;
in addition, the almost full moon (full moon Dec 16) is reported to 
have a halo (ring or arc) around it;
apparently, some part of the lunar halo display (e.g. an arc, called here {\em white qi} or vapour) 
was located in Taurus, and it followed the motion of the moon during and after the occultation 
of the Pleiades; the whole event started in the east and ended in the south, 
which is not possible for aurorae.
\item AD 763 Nov, China: 
{\em In the first year of the Daizong Emperor, Tibetan forces invaded the Bian Gate Bridge.  
The Emperor visited Shan County. The royal armies found themselves at a disadvantage, 
as there was constantly a purple qi like the canopy of a chariot that met their horses' heads.  
Once they returned to Tong Pass, the Emperor sighed and said: 
How broad and vast are the waters that send me off to the east ...} 
from DC80 from Su E's 9th Duyang zabian 
(translated from Chinese by J. Chapman, priv. comm.).
This event was misinterpreted as aurora in DC80 
and also listed as aurora in S98 for AD 763 Nov (for Tongguan, China),
but neither in YSW nor in XPJ.
There is no direct indication that the purple {\em qi} here were seen in the sky. 
The {\em qi} was reported to be constant and on the same level as the horses.
It was most likely just fog.
\item AD 767 Aug 25, Xian, Shaanxi, China: {\em At 17-19h o'clo\-ck, 
there was white vapour spreading over the sky} from K73
from Chiu-t'ang-shu, classified by K73 as {\em probable to doubtful}; 
listed as aurora in U13, but neither in YSW nor in XPJ; new moon was on Aug 28/29,
probably not an aurora, because the {\em white vapour} was seen around sunset (18:30h),
civil twilight ended at 18:56h local time for this place and time,
so that the end of the 17-19h double-hour is during nautical twilight
(astronomical twilight ended 19:58h). perhaps, it was a solar halo display around sunset.
\item AD 767 (between Jul 30 and Aug 28, probably Aug 25), 
Korea: {\em The Emporer attended dinner at the Hall of State Ceremonies,
and he saw three stars drop down the garden of the Palace to struggle with one another. Their lights
so brilliant as if fire spouting and dispersing} from K73 from Samguksagui and Shillapongui for Korea;
{\em doubtful} according to K73; listed as aurora in U13, but neither in YSW nor in XPJ;
the description indicates meteors (probably the Perseids).
K73 suggested Aug 25 as probable date, most likely just because the previously listed
(but different) event in Xian, China, was dated to AD 767 Aug 25.
\item AD 773 Aug 9, Xian, Shaanxi, China: 
{\em At 17-19h o'clo\-ck, there were three bands of vapour briged over the sky},
from K73 from Chiu-t'ang-shu, {\em probable to doubtful} according to K73;
it cannot be an aurora, because it happened in the 17-19h double-hour (end of civil twilight at 19:16h
and sunset at 18:49h for this time and place); hence, maybe a solar halo display with three segments;
full moon was on Aug 6/7, so that the moon rose after the 17-19h double-hour; 
listed as aurora in U13, but neither in YSW (as already noticed by Cliver et al. 2014) nor in XPJ.
\item AD 773, England: misinterpreted and misdated ({\em red crucifix/cross}),
e.g. by U13,
H96 (citing Johnson (1880), who took it from the Anglo-Saxon Chronicle), 
and S98: see AD 776 below.
\item AD 773, Fritzlar, Germany: {\em apparition interpreted by Christians as riders 
on white horses} misdated to AD 773 in U13, see next item.
\item AD 774, Fritzlar, Germany: {\em apparition interpreted by Christians as riders on white horses} 
as given by U13 for AD 773 from Scholz \& Rogers (1970),
who translated the apparition as {\em two young men on white horses};
the event is narrated in the Royal Frankish Annals,
according to the context after Christmas AD 773, so that it was certainly AD 774;
the Latin text ({\em duo iuvenes in albis}) can mean {\em two young men on white horses} or {\em in white}.
Our interpretation of the sighting as mock suns is supported by the frequent use
of the antique {\em dioscuri motive} adapted by Christians in medieval times
(see Neuh\"auser \& Neuh\"auser (2014) for a more detailed discussion). 
\item AD 774, England: misinterpreted and misdated ({\em red crucifix/cross}),
e.g. by U13,
H96 (citing D79, who did not list the {\em red cross}, 
except by quoting L62 ({\em 774 (L)}), but D79 did not give the text),
and S98: see AD 776 below.
\item AD 774 Dec 11, China: misinterpreted and misdated by Zhou et al. (2014): see AD 776 Jan 12/13 below.
\item AD 775 Jan 17, China: misinterpreted and misdated by Zhou et al. (2014): see AD 776 Jan 12/13 below.
\item AD 775 Dec 31, China: misinterpreted and misdated by U13 and S98: see AD 776 Jan 12/13 below.
\item AD 775, Mechelen, Belgium:
{\em Celestial lights marked the place where it lay [the body of Irish saint Rumoldus, death reported for
AD 775 June 24, funeral soon later] and ... a mysterious flame of light was seen ... attracted the
notice of fishermen ... Night after night the light appeared} published by McClintock \& Strong (1880).
This report was presented as aurora in Zhou et al. (2014) with the following text:
{\em In Italy, in AD 775, the celestial lights as recorded in an Encyclopedia (McClintock \& Strong 1880)}.
This text was written or compiled by Theodoricus (prior of Saint-Truiden abbey) around AD 1100.
Apart from the wrong location given in Zhou et al. (2014), 
there are strong doubts on the dating, because that area was christianized several decades earlier,
so that Rumoldus was killed (a holy martyr for the christians) at that time,
and the $^{14}$C dating of 40 bones of his remains indicates 
a death date between AD 580 and 655 (Strydonck et al. 2009).
Hence, we cannot consider this report in our aurora catalog given the dating.
\item AD 776 (presumably 774), England: 
The report about the red cross ({\em This year also appeared in the heavens a red crucifix, after sunset},
Allen 2012)
and reports derived from it were misinterpreted
once as absorbed supernova (Allen 2012) and several times as aurora 
(Sch\"o\-ning 1760, J70, L62, 
H96,\footnote{H96 cites the Anglo-Saxon Chronicle text and gives {\em ASC 50} as reference, but {\em ASC} is given
as a different paper by Pang \& Bangert (1993) about an unrelated planetary conjunction BC.}
S98, Schlegel \& Schlegel 2011,
U13, etc.);
different translations give either cross or crucifix and then either sky or heaven(s).
L62 and U13 combined the {\em red cross/crucifix} from the Anglo-Saxon Chronicle
or its reception by others with a report about {\em serpents in South-Saxon},
also from the Anglo-Saxon Chronicle, interpreted as indicating motion on sky.
While the {\em red} colour given to be observed {\em after sunset} (and presumable motion)
might fulfill some aurora criteria: red color, motion (serpents ?), and night-time (after sunset ?).
However, the timing {\em after sunset} usually means twilight;
in medieval English (and Latin) the word {\em after} or \ae{\em fter} (in Latin: {\em post}) can also
mean {\em during} (sunset). The serpents are actually adders --
the Anglo-Saxon Chronicle does not connect the adders with the {\em red cross/crucifix}, 
but mentions a {\em battle a Otford} in between, which took place in AD 776 (Stenton 1970).
There is a better explanation for the red cross observation:
If during sunset, then it was a solar halo display at least with horizontal arc and vertical pillar, 
together apparently looking like a red cross during sunset
-- or, if truly {\em after} sunset, then it was a lunar halo display
(see e.g. N72, Minnaert 1993).
See Neuh\"auser \& Neuh\"auser (2014) for a very detailed discussion.
\item AD 776 Jan 12/13, Xian, Shaanxi, China: {\em At night, abo\-ve the moon in the east,
there were more than ten bands of white vapour like a glossed silk,
penetrating the We-ch (Aur), Tung-ching (Gem), Yu-kuei (Cnc), Tsuei (Ori), Shen (Ori), Pi (Tau),
Liu (Hya), and Hsien-yuan (Lyn, Leo, or LMi). Just after midnight, they vanished} from K73
from Chiu-t'ang-shu, {\em very probable} according to K73,
dated by him to AD 775 Dec 31 ({\em jiazi day}, i.e. 1st day of the sexagenary cycle) 
based on the text variant in Liu (2007), also in U13, but neither in YSW 
(as already noticed in Cliver et al. 2014) nor in XPJ. 
Since new moon was on AD 775 Dec 26/27, 
the 4-day-old moon could not be seen {\em in the east} on
AD 775 Dec 31. Another text variant (Liu 1975) gives {\em bingzi day} (13th day of the sexagenary cycle)
{\em after the third watch} (22:53h to 01:25h local time at that place on that date), 
hence AD 776 Jan 12/13 (J. Chapman \& M. Csikszentmihalyi, priv. comm.) 
consistent with lunar phase: Full moon was on AD 776 Jan 10/11, so that the moon
was in the east in the first half of the night (moon rise at Xian in Shaanxi around 19:40h local time).
All constellations mentioned rise in the 1st half of the night from the E and move towards S with the moon 
(in Leo that night), which is not consistent with aurorae.
The {\em bands of white vapour like a glossed silk} ({\em above the moon}) may have been a lunar halo effect.
The {\em moon} is otherwise never mentioned close together with true aurorae in Chinese reports,
{\em silk} is only very rarely mentioned (as checked and confirmed for AD 550 to 845). 
Note that the bands of white vapour were mentioned to be {\em only} above the moon.
(Aurorae are possible during full moon nights, as shown for AD 796 Oct 20 (Sect. 3.2), 
but only in the north, not in the direction of or close to the moon.) \\
This particular report is probably listed twice in Cliver et al. (2014), 
once for the end of AD 775 and once for early AD 776:
{\em Usoskin et al. (2013) noted that the 775 event was 
accompanied by a cluster of low-latitude aurora reported from Shanxi\footnote{The transcription
of the Chinese {\em Shaanxi} province as {\em Shanxi} in U13
and K73 is misleading, as there is another province
called {\em Shanxi}, while the old capital (now Xian) is located in the province called {\em Shaanxi}.} Province, 
China, in 770 (twice), 773, and 775 based on the
compilations of Keimatsu (1973) and Yau et al. (1995) ... 
From a re-examination of Keimatsu (1973), F.R. Stephenson
(2013, priv. comm.) ... suggests ... that a report of more than 
10 bands of white vapour on AD 776 Jan 12 may
be auroral in nature\footnote{We would like to note that
Yau, Stephenson, and Willis (1995) did not classify this event as aurora.}. 
In any case, the evidence for a grouping of strong 
auroral activity ca. 775 is weakened} (Cliver et al. 2014).
With the report for AD 775, almost certainly both U13 and Cliver et al. (2014) mean the report with the
wrong dating based on the text variant in Liu (2007), even though it is inconsistent with the moon phase,
while for the AD 776 event, Cliver et al. (2014) most certainly mean the same event, but based
on the text variant by Liu (1975), which is consistent with the lunar phase. \\
Zhou et al. (2014) list this particular event as aurora for {\em 11 Dec 774, i.e. 17 Jan AD775}; 
apart from the fact that we do not understand the two dates (both wrong), 
the almost full moon on the true date (AD 776 Jan 12/13, 22:53h to 01:25h) was visible since about 19:40h local time,
so that the sighting did not start as early as 17-18h as mentioned in Zhou et al. (2014);
since the observed phenomenon was seen {\em above the moon} (almost full moon) in south-eastern direction, 
it was almost certainly not an aurora. \\
Zhou et al. (2014) then list two other presumable aurorae, 
both misdated for AD 774/5 (see above: 
AD 774/5 {\em red cross} and {\em celestial lights} at Mechelen), 
neither of which is a true aurora, 
and they claim that they show evidence for worldwide strong aurora activity. 
This claim is not substantiated.
Zhou et al. (2014) argue that {\em the date difference may be from the inaccurate records 
for the dates of auroras on Europe}. 
We would like to point out, however, that all three events were misdated by Zhou et al. (2014) 
by at least one year, and in one case, maybe more than a century -- in addition, 
at least two of the three reports do not relate to aurorae. 
The one event they cite that might have been an aurora ({\em celestial lights}) 
would have been a very weak one without colour, but was several decades earlier. \\
Zhou et al. (2014) conclude: 
{\em The strongest AD775 auroras in the past 11400 years were first successfully 
identified with the historical records ..., 
the super auroras were generated in January AD775 and lasted about 8h.} \\
This claim is not supported by the historical records,
the arguments by Zhou et al. (2014) are not convincing:
Zhou et al. (2014) listed for AD 707 another aurora with red colour for three days
(and there in another similar report for AD 708, both in Yau et al. (1995), but not in Xu et al. 2000), 
which they should have considered much stronger than the presumable aurora around AD 774/5, 
but there is no additional (or even stronger) $^{14}$C peak around AD 707.
For a discussion of a report about some {\em white vapour}, presumably in connection with a comet
misdated to AD 776 January, and its possible relation to the {\em white vapour} discussed here,
see Chapman et al. (to be submitted).
\item AD 776, Germany: {\em two shields burning with red colour and moving above the church itself} as cited by
Gibbons \& Werner (2012) from the Annales Laurissenses (maio\-res), the chronicle of monastery Lorsch in Germany,
it is basically identical to the Royal Frankish Annals.
This wording is quoted as {\em the likeness of two shields red with flame wheeling over the church} 
in the English translation by Scholz \& Rogers (1970), cited by U13 
($^{\prime \prime}${\em inflam\-ed shields$^{\prime \prime}$ in the sky}) from L62 in Latin for Germany
and also by S98 for AD 776 for both Germany and England, 
and interpreted as aurora by all of them.
Obviously, this happened during the day (as noticed by Gibbons \& Werner 2012), 
as mentioned explicitly ({\em on a day} or {\em quadam die}).
In medieval and ancient times, mock suns were often interpreted as theophania
(intervention by god), as it becomes clear here from the context.
An interpretation of the {\em two shields} as red mock suns 
is much more likely (see also Schreiber 1984).
See Neuh\"auser \& Neuh\"auser (2014) for a more detailed discussion.
\item AD 778, Central Europe: {\em On 31 January, 
two war arm\-ies appeared to fight on sky} (P88 from Lycosthenes),
also given in M33, F73, and L62 for AD 778 Jan 31 (all quoting Lycosthenes). 
However, as shown by L62, the text in
Lycosthenes speaks about an eclipse, which was not visible in Europe in AD 778,
then about the {\em war armies}, and then about a Mercury transit accross the Sun (which also
did not happen in AD 778, neither a Venus transit); since a Mercury transit was reported in the
Royal Frankish Chronicle for AD 807 (which was probably a sunspot) together with an aurora a few weeks earlier 
and three eclipses visible in Europe within about one year, this report by Lycosthenes 
belongs to AD 807. Lycosthenes often has dating errors (e.g. L62);
Lycosthenes has this text in different variants in AD 778, 803, and 808, correct is only AD 807. 
\item 786 Dec 19, France: {\em six days before Christmas, 
thunder and blizzard in all of France ... and a bow on the sky
in the clouds appeared at night} from L62 from Royal Frankish Annals
which was classified as aurora in L62,
H96,
and S98, 
but which was likely a thunderstorm at night plus a bow in a (different ?) night;
while the {\em bow ... in the clouds} (see Gen 9, 13+14) might well be a night rainbow or a lunar halo
(reported after Dec 19 ?), new moon was on Dec 25, so that it was not bright enough for a
rainbow or halo in the rest of the month after Dec 19;
while the {\em bow} could have been observed before the {\em thunder} around full moon,
it is also not impossible that this {\em bow} means an aurora element 
(see aurora in AD 786 Dec, Sect. 3.2).
L62 combined these two events with another source about the true aurora in AD 786 Dec.
\item AD 793, York, England: {\em What did the flux} [pluvia] {\em of blood mean that}
[MS: {\em qui}, but should be {\em quam}]
{\em we saw during the 40-day-period} [Lent] {\em in the city of York at the church of Saint Peter,
the founder of the apostels, which is the main seat of the whole empire,
as it was descending from the northern side} [{\em de borealibus}] 
{\em of the building during clear sky} [{\em sereno aere}] {\em from the peak of the roof?
How can it not be thought that a penalty of/with blood from the north} [{\em a borealibus}] 
{\em will come upon our people?} \\
(Alcuin Letter no. 16 as translated by us from Latin as found on eMGH\footnote{www.dmgh.de/de/fs1/object/display/bsb00000538$\_$00051.html});
this event was interpreted as aurora by Schove (1983, 1984), who gave {\em rain of blood}; 
Hetherington (1996) cited {\em a $^{\prime}$shower of blood$^{\prime}$ seen to the north from York
on a fine night} following Schove (1983), but the timing ({\em night}) is not justified by the Latin text.
Dutton (2008) translated the Latin {\em sereno aere} with {\em on a clear day}
and interpreted it as Sahara dust, but the Latin wording does not specify whether it was day or night;
Sahara dust would be expected from the south, not from the north;
the text quite clearly specifies that something like a {\em flux of blood} 
descended down {\em from the peak of the roof} {\em from the northern side
of the building}, so that it would be too speculative to consider it an aurora seen on the sky behind the church;
also, we have never read about a {\em clear sky} during an historic aurora sighting; 
also given the clear sky, it was not real rain,
but maybe reddish liquid drops from many butterflies leaving their cocoons,
or blood drops from many bees leaving their beehive (see e.g. Schwegler 2002),
or other insects or birds or something similar (located under the roof), 
which may already be possible in an early warm spring.
The {\em penalty of/with blood from the north} is not a portent (like a sign or aurora),
but means the attack of the Vikings, who came from the north(-north-east).
Alcuin was an advisor to Charlemagne; while the latter started the year at Easter until AD 800, it may not
be fully clear as to which event or day 
was considered as start of the calendar year in England or by Alcuin around AD 793 --
the Anglo-Saxon Chronicle at around this time used different definitions; however, since
Alcuin, in this letter, considers the sighting as one of several portents 
for the first attack of the Vikings to England (AD 793 June 8),
it is quite likely that the sighting happened in Lent of what we consider AD 793;
the {\em 40-day-period} clearly refers to Lent, the 40 days before Easter (first full moon in spring of 793 was
on Monday Apr 1, so that Easter was on Sunday Apr 7), i.e. Lent was mostly in March 793 
(eMGH give AD 793 Feb 20 to Mar 31 as Lent).
\item AD 803: See AD 778.
\item AD 808: See AD 778.
\item AD 817 Oct 26, Xian, Shaanxi, China: {\em At 19-21h o'clo\-ck, 
there was a meteor shooting from the midheaven ...} from
K74 from T'ang-shu, classifying it as {\em probable}, but obviously 
a meteor in the evening (full moon Oct 28/29), maybe from the Orionids,
listed three times in S98, but neither in YSW nor in XPJ.
\end{itemize}

We omit all those events in our statistics.

\begin{figure*}
\begin{center}
{\includegraphics[angle=270,width=15cm]{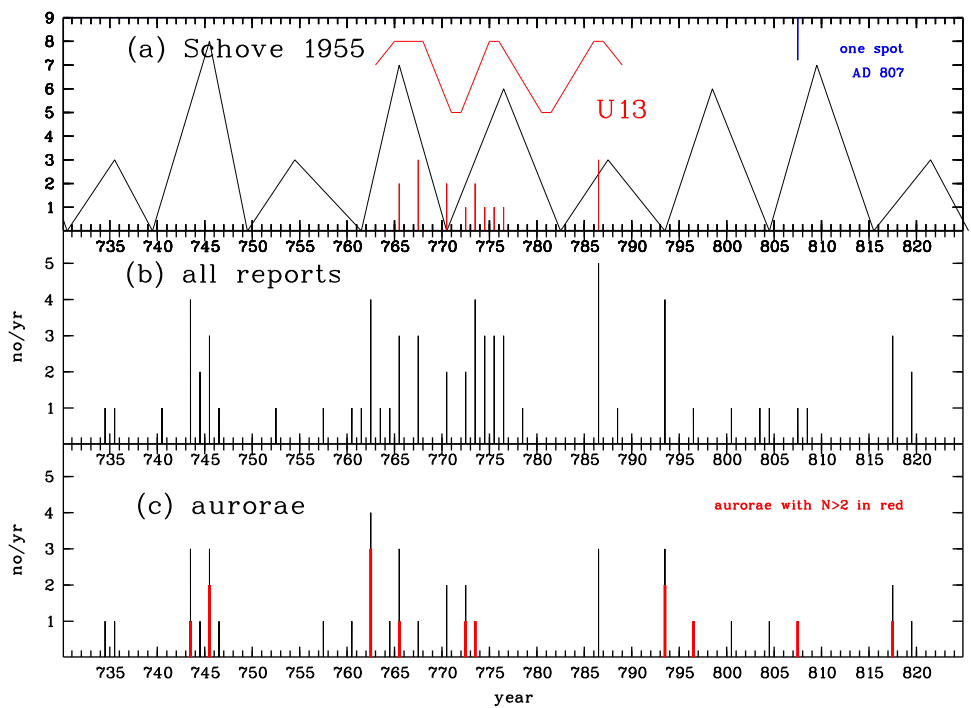}}
\caption{{\bf Aurora reports and solar activity reconstructions for the time AD 731 to 825:}
(a) Top: Schwabe cycle reconstructions from Schove (1955) in black and,
for a shorter period, Usoskin et al. (2013) (U13) in red together
with the reports interpreted as aurorae by U13 as red lines -- and the one sunspot in AD 807
in blue. The strengths of the solar activity maxima identified by Schove (1955) are
indicated by the scale on the left hand-side axis (1 for the weakest to 9 for the strongest).
(b) Middle: All reports suggested so far as aurorae, 
either the likely true and potential
aurorae (our Sect. 3.2 including four events with N=0, as potential aurorae)
or misdated or misinterpreted events (our Sect. 3.4).
(c) Bottom: Only those events listed by us as likely true aurorae 
(our Sect. 3.2, without potential aurorae with N=0).
Those events where the number of aurora criteria N fulfilled is N=1-2 are plotted as black.
Events with N being larger than 2 are indicated 
as thick (red) lines; 
those are more likely and probably also stronger events.
In particular, the misinterpreted events in the middle of the AD 770s
led U13 and Schove (1955) to conclude on a solar activity maximum in the mid AD 770s.
We plotted the reconstructed minima and maxima
as well as the individual events into the middle of the respective years.}
\end{center}
\end{figure*}

\subsection{Comparison with Silverman's and Usoskin's lists}

We discuss here the similarities and differences between our aurora catalog for
AD 731 to 825 and the aurora catalogs from Silverman (1998) and Usoskin et al. (2013).
We list a total of 39 likely true aurorae 
(plus four potential aurorae with N=0) 
in Sect. 3.2; in this count, the aurora reported from Yemen for AD 816/817
is not counted, because it is more likely a copy of
an aurora from Iraq for AD 817); the events in AD 793 May in Iraq are listed as two aurorae, 
as explained in Sect. 3.2, because they lasted for more than four days; 
the event(s) in AD 743 June
are listed (in Sect. 3.2) and counted twice, as they were reported independently 
(with complementary information) in two different areas.

Silverman (1998) has 49 entries from AD 731 to 825 in his online aurora catalog\footnote{nssdcftp.gsfc.nasa.gov/miscellaneous/aurora},
but some events are listed several times, and some of them are likely not true aurorae: \\
The aurorae for AD 757, 760, 762 (Europe), 764, 765 (two) are listed as aurora (once each) by Silverman, \\
the aurora from AD 745 Jan 1 in England 
is listed five times (for AD 743 to 748 in England and Middle East), \\
the aurora from the Anglo-Saxon Chronicle for AD 793 (England) is listed four times, 
(namely three times for 793 and once for 794, referencing publications about the Anglo-Saxon Chronicle aurora;
the AD 794 event is without exact date, by citing {\em Lowe (1870)}, 
given there as {\em brilliant}, this is probably again the same from the Anglo-Saxon Chronicle), \\
the aurora AD 762 May 1 is listed four times (including once as aurora in AD 762 Aug
from Dai \& Chen (1980), \\
the (potential) aurora AD 740 (Germany), 
and the aurorae in
762 May 20, 762 Sep 16, 786, and 796 
(all China) are listed three times each, \\
for the two aurorae listed for 742/743 (presumably from Italy), 
we list the two aurorae for AD 743 for Constantinople and Near East, \\ 
the aurorae listed for AD 807 and 817 (Germany) are listed twice each, \\
the event listed for AD 775 Dec 31 was in fact on AD 776 Jan 12/13 (Sect. 3.4), 
but {\em above the} almost full {\em moon}, hence not an aurora, but a lunar halo effect, \\
the events in AD 761 (China), 763 (China), 776 (Germany), 786 (France), and 817 (China) 
are listed in Sect. 3.4 as different (non-auroral) events, \\
the events listed for AD 773, 774, and 776 (Great Britain) are all the {\em red cross after sunset}, 
which was a halo cross. \\
Hence, there are 17 different likely true aurorae 
(plus one potential aurora with N=0, AD 740) remaining in Silverman's catalog.
Our catalog of likely true aurorae (Sect. 3.2) includes 39 aurorae 
(including all likely true aurorae in Silverman 1998),
plus four potential aurorae with N=0.

U13 listed 14 to 16 different events\footnote{With the wording
{\em the next nearest [oriental] observations at AD 767}, 
U13 could mean any (i.e. one or two or three) of the
three events listed in Keimatsu (1973) for AD 767,
hence in total 14 {\em to} 16 events.} as aurorae from AD 765 to 786, 
but some phenomena listed are halos or otherwise doubtful: \\
Two East Asian events listed for AD 767 Aug 25 (day-time) and AD 767 Jul/Aug
were meteors sighted in Korea (see previous footnote), \\
the Chinese event listed for AD 773 Aug was from 5-7h pm, hence during the day, i.e. not an aurora, \\
the Chinese event listed for AD 775 Dec was in 776 Jan {\em above the moon}, hence not an aurora
(see also Chapman et al. to be submitted), \\
the European events listed for AD 773 (two young men on white horses), 774 (red cross), and 776 (two
inflamed shields) were all halo phenonena (see Neuh\"auser \& Neuh\"auser 2014). \\
Of the 14 to 16 events listed, up to nine are likely true or potential aurorae 
(including one in AD 786 with N=0, i.e. potential only: 
{\em A horrible vision in Cluain Moccu Nois, and great penance done throughout Ireland}). 
Our catalog of likely true aurorae (for the period studied in U13, AD 765 to 786) 
includes 12 likely true aurorae plus one event with N=0 (including some events not listed in U13).

In Fig. 2, we compare the distribution of the likely true aurorae (Sect. 3.2)
with all events (including the previously misinterpreted and/or misdated events, Sect. 3.4).

\section{Schwabe cycles from AD 731 to 825 (844)}

Given the historic observations of sunspots (Sect. 2) and aurorae (Sect. 3.2)
as well as the $^{14}$C record recently published with 1- to 2-yr time resolution
(M12, U13, Miyake et al. 2013a, Jull et al. 2014), 
we can now try to reconstruct the solar Schwabe cycles for the period studied.

There have been two such attempts previously: Schove (1955) used sunspots and aurorae
of the last millenia -- assuming that there are always nine solar activity maxima per century and that
the time between two successive maxima was 8-16 yr. 
More recently, U13 used aurorae only for the reconstruction
of three Schwabe cycle activity (aurora) maxima around AD 775. 
Schove (1955) gave AD 765 as very strong maximum, AD 776 as moderately strong, 
but uncertain maximum (given by him in brackets),
then AD 787 as weak uncertain maximum given in brackets. 
He also listed AD 770, 782, and 793 as uncertain minima.
U13 claimed to have found a {\em high solar activity level} around AD 775
plus {\em the next nearest (aurora) observations around AD 765-767 and AD 786}.

Miyake et al. (2013b) used their $^{14}$C record with 2-yr time resolution 
(Miyake et al. 2013a) to find the mean cycle length in the period AD 600 to 710. 
They gave 10 yr as mean period length for AD 600-630 and 12-13 yr for AD 630-710 (Dark Age Grand Minimum), 
but did not specify the times of activity or $^{14}$C maxima and minima.
If the period length would indeed be 12-13 yr from exactly AD 630-710,
then 6.2-6.7 cycles fit into this time interval.
The time interval studied by Miyake et al. (2013b) for Schwabe cycle lengths (until AD 710)
does not overlap with the period studied by us (since AD 731).

We should keep in mind that aurora peaks can happen at different phases of
the sunspot Schwabe cycle, namely just (one to few years) before and after the sunspot maximum
and also in the declining part (Sect. 3.1).
The strong\-est flares of the last two centuries happened
in the middle of or early in their Schwabe cycle, e.g. the Carrington event at the end of August 1859
occurred in the 4th year of that Schwabe cycle and 7.5 yr before the next minimum
(minima in AD 1856.0 and 1867.2 according to Hoyt \& Schatten 1998), and
the strongest flare in the last century in February 1956 happened 2 yr
after the minimum in AD 1954.3. 
Hence, it may be difficult to reconstruct Schwabe cycle phases from aurorae only.

While the two previous attempts to reconstruct Schwabe cycles 
used aurorae and sunspots (Schove 1955, U13),
we will also use the $^{14}$C record,
which is now available with 1- to 2-yr time resolution 
(M12, U13, Miyake et al. 2013a, Jull et al. 2014).

We show in Fig. 1 all proxies for reconstructing solar Schwabe cycle maxima and minima:
$^{14}$C as well as sunspots and aurorae.
$^{14}$C is incorporated in tree rings within the first $\sim 2-3$ years (Sigman \& Boyle 2000),
$^{10}$Be is incorporated mostly within the first year (Heikkil\"a et al. 2008). 
We plot the $^{14}$C data 2.5 yr ahead of its direct measurement date,
as recommended by Houtermans et al. (1973), Braziunas (1990), and Stuiver (1994);
given the carbon cycle, 2-3 yr after its production, there is the largest yearly amount of incorporation 
into tree rings; nevertheless, there is also a significant incorporation
in the very first year.\footnote{The trees used in this study are all on the northern hemisphere, 
so that the incorporation of carbon into tree rings happens mostly around the middle
of the calendar year (e.g. for year $x$ at epoch $x.5$), so that we shift from the
measurement (or incorporation) date $x.5$ to $x-2$, e.g. the $^{14}$C measured
in the AD 774 tree ring was incorporated at 774.5 and is plotted at 772.0. 
Those 2.5 yr do include the Forbush delay.}
Hence, $^{14}$C data shifted back by 2.5 yr are to be considered a good approximation
of the $^{14}$C production rate, which we do not use to avoid further assumptions.
$^{10}$Be data at this time have a much lower time-resolution, 
so that they cannot be used here for Schwabe cycles, but $^{10}$Be is also plotted in Fig. 1.

Solar activity (wind) modulates incoming cosmic rays, 
so that we expect an inverse Schwabe cycle in radioisotopes:
Around an activity maximum, we expect a minimum in $^{14}$C (shifted by 2.5 yr) and viceversa. 
This is indeed seen in Fig. 1:
We show by dotted lines in the first half of the 9th century 
(clusters of) sunspots and aurorae, which are consistent with
local minima of $^{14}$C.
When we indicate that {\em solar activity} is {\em maximal}, 
then we mean that the activity phase is at least not at the minimum -- 
we then have low levels of $^{14}$C and high levels of aurorae and sunspots. 
{\em Activity minima} are obtained from local $^{14}$C maxima
($^{14}$C mostly has 2 yr time resolution only).
The arrows in Fig. 1 roughly mark the activity minima of the Schwabe cycles. 

We list below the Schwabe cycle phases, where activity is minimal and maximal 
(we consider here only very likely true aurorae with N $\ge 3$
and spots, if available).
Radiocarbon dates given are {\em shifted} by those 2.5 yr 
(unless otherwise noted).\footnote{The error bars or time ranges given do not include the
additional error bar from the carbon cycle and the Forbush effect (we shift by exactly 2.5 yr);
in alternating A+/A- cycles, the Forbush delay is either a few months or about one yr.} 
\begin{itemize}
\item Activity minimal at AD $802 (\pm 1)$ as a local $^{14}$C maximum in data with 2 year time resolution
(hence, the error bar $\pm 1$ giving a range of 2 yr)
after having been shifted (like all $^{14}$C data) by those 2.5 yr due 
to the carbon cycle to consider the $^{14}$C production rate.
(There are also no aurorae in AD 801, 802, and 803.)
\item Activity maximal around AD 807 with one sunspot at AD 807 Mar and one strong aurora at AD 807 Feb
together with a broad local $^{14}$C minimum $\sim$ AD 804 to 810;
if the spot observation from Europe for AD 807 Mar ({\em spot a little above the center}
of the Sun) indicates a spot at low heliographic latitude, the Schwabe cycle
is certainly not at its start in AD 807, but may well be in its maximum phase or second half.
\item Activity minimal at AD $812 (\pm 1)$ to $814 (\pm 1)$ as a local $^{14}$C maximum.
(There are no aurorae from AD 808 to 816.)
\item Activity maximal at AD 817-820 without sunspots, but with aurorae in Germany and Iraq 
(quasi-simultaneous) in AD 817 and a local $^{14}$C minimum $\sim$ AD 818 to 820.
\item Activity minimal at AD $822 (\pm 1)$ to $824 (\pm 1)$ as a local $^{14}$C maximum.
(There are no aurorae from AD 820 to 825.)
\item Activity maximal at AD 826-832 with sunspots in AD 826, 829, and 832, 
strong aurorae in AD 826 and 827,\footnote{These aurorae are not given in Sect. 3.2, where
the listing ends in AD 825; details can be found in the publications listed in Sect 3.2; 
they will be included in our extended aurora catalog from AD 550 to 845 (in prep.).}
and a broad local $^{14}$C minimum $\sim$ AD 826 to 832. 
\item Activity minimal at AD $834 (\pm 1)$ as a local $^{14}$C maximum.
(There are no aurorae from AD 829 to 835.)
\item Activity maximal at AD 836-841 with sunspots in AD 837, 840, and 841, 
many strong aurorae in AD 839 \& 840 (see previous footnote),
and two local $^{14}$C minima in $\sim$ AD 836 and 840 (note that solar activity maxima are often
related to two local aurora maxima, so that two local $^{14}$C minima are not surprising);
we have additional aurorae in AD 842 and 843, apparently the declining phase of the Schwabe cycle.
If the spot observation from Arabia for AD 840  
({\em a black spot close to the middle of the Sun}) 
indicates a spot at low heliographic latitude, the Schwabe cycle
is certainly not at its start in AD 840, but may well be in its maximum phase or in its second half.
\item Activity minimal at AD $844 (\pm 1)$ as a local $^{14}$C maximum.
\end{itemize}

Then, we find two more phases, where solar activity is maximal, even without further sunspots:
\begin{itemize}
\item (Activity minimal at AD $802 (\pm 1)$, see above)
\item Activity is maximal at AD 796-800 with a strong Chinese aurora in AD 796
and a local $^{14}$C minimum AD 796 to 800 (in Miyake et al. 2013a data).
\item Activity minimal as a local $^{14}$C maximum at AD 794 $(\pm 1)$ or $\sim 794$ considering 
the Jull et al. (2014) Si\-be\-rian tree with largest $^{14}$C amplitude and 1-yr time resolution.
The aurorae in AD 793 (one in England, one or two in Iraq) 
are either in the declining phase of the previous cycle or,
more likely, at the start of the new cycle (no aurorae from AD 787 to 792) 
-- strong aurorae shortly after the activity minimum are not exceptional.
\item Activity maximal at AD 786-790 
with a local $^{14}$C minimum from (at least) AD 787 to 790, where the $^{14}$C data from three 
different trees agree with each other (Miyake et al. 2013a and two trees from Jull et al. 2014)
with two aurorae with N=2 in AD 786 from China and Europe (quasi-simultaneous).
\item Activity minimal at AD $784 \pm 1$ as a local $^{14}$C maximum considering all
three data sets (Miyake et al. 2013a, Jull et al. 2014 for two trees).
There were no aurorae from AD 774 to 785.
\end{itemize}

The mean length (from maximal to maximal activity as listed above) 
identified above between about AD 788 and 838.5 is $10.5 \pm 0.8$ yr 
(from six maxima with five time periods inbetween).
From the first activity minimum (about AD 784)  
until the last activity minimum (about AD 844) as listed above,
there are 60 yr for six Schwabe cycles with a mean length of $10.0 \pm 0.9$ yr.

For the time before the AD 775 $^{14}$C variation, 
we can identify activity maxima and minima as follows
(since AD 733 $^{14}$C decreases in both Intcal and Miyake et al. (2013a), 
which should be due to an activity increase):
\begin{itemize}
\item Activity minimal at AD 732-734 $(\pm 1)$ as local $^{14}$C maximum 
(in Miyake et al. 2013a data with 2-yr time resolution).
(There were no aurorae in AD 731-733.)
\item Activity maximal around AD $736 (\pm 1)$ and $740 (\pm 1)$, where we see local $^{14}$C minima.
(There were aurorae with N=2 in AD 734 and 735.)
\item Activity minimal at AD $742 \pm 1$ as a local $^{14}$C maximum; 
since $^{14}$C is decreasing, the small local $^{14}$C maximum is hard to recognize,
so that the activity minimum may be at AD $742 \pm 1$;
from AD 743 to 748 (as plotted) Intcal09 data have one of the largest decreases in a 5-yr time step
(similar in Intcal13, Reimer et al. 2013),
which is not seen in Miyake et al. (2013a) data, 
but it is consistent with strong aurorae AD 743 and 745, the latter being simultaneous
in England and Near East. 
\item Activity maximal with reliable aurorae in AD 743 and 745 (partly in the Near East), 
a local $^{14}$C minimum apparently late AD 752-754;
but, as mentioned just before, the Miyake et al. (2013a) data from AD 744 to 750 do
not reflect the strongly decreasing Intcal09 data, 
so that the available $^{14}$C data (2 yr time resolution) may be less credible here.
We can clearly recognize two local $^{14}$C maxima, one in AD 732-734 $(\pm 1)$ 
and then one in AD 756-758 $(\pm 1)$,
so that two Schwabe cycles fit into this period.
\item Activity minimal around AD 756-758 $(\pm 1)$ as local $^{14}$C maximum.
(There are no aurorae until AD 756.)
\end{itemize}

The previously mentioned secular $^{14}$C decrease reaches its lowest level 
after AD 757 and remains low in the AD 760s and the early AD 770s
(with 1-yr time resolution in three data sets, namely U13 plus two trees in Jull et al. 2014,
and since AD 768 (shifted time) also in M12 for two trees).
We also see an aurora cluster from AD 757 until 773,
in particular very credible aurorae in AD 762 (several), 765, 772, and 773
from the Near East and, for the first time in our study period, also from China
(lowest geomagnetic latitude), indicating very strong activity. 
The long aurora cluster obviously consists of two rather strong Schwabe cycles 
(several very low $^{14}$C minima).
A short activity minimum inbetween the two cycle maxima appears to be around AD 766
(local $^{14}$C maximum), which is not very pronounced.

The next local maximum in $^{14}$C appears in AD 775-777 in the unshifted data sets
(and remains similar high until AD 777-779 in the different data sets).
After shifting by the usual 2.5 yr, the maximum is AD $\sim 774 \pm 1$.
The Schwabe cycle activity minimum should lie around that time.
Those four more Schwabe cycles from the activity minima around AD 733 to around AD 774
have a mean length of $10.3 \pm 3.2$ yr.

The last two cycles together from around AD 757 to about AD 774 lasted $\sim 17$ yr only,
they were very strong cycles.
Just before the start of the short Dalton minimum, there were three very strong
Schwabe cycles (no. 2, 3, 4), which had lengths of 8.8-9.0 yr only,\footnote{considering 
that the very long cycle no. 4 has to be split into two cycles, the first of which would be 8.8 yr}
namely with
minima in AD 1766.5, 1775.5, 1784.3, and 1793.1 (Hoyt \& Schatten 1998, Usoskin et al. 2001).
The activity minima inbetween these strong cycles before the Dalton minimum
were not deep and have shown moderate activity (e.g., Hoyt \& Schatten 1998).
Therefore, it is not surprising to have aurorae in AD 765 -- in or near the activity minimum
between those two strong cycles before AD 775.
Activity increased in the four decades before AD 775, so that the $^{14}$C level decreased --
similar in the four cycles before Dalton.

Towards the end of the Schwabe cycle, which started around AD 766, we have the following situation:
The unshifted $^{14}$C data in AD 774 are already higher than in AD 773 showing decreasing 
(but still relatively high) solar wind (i.e. shifted in AD 772/1);
there were two aurorae in the records for AD 772 (one with N=3 in the summer in Amida, Syria/Turkey).
Radiocarbon continued to increase from AD 774.5 to 775.5 (unshifted) or AD 772 to 773 (shifted);
there is only one aurora known for AD 773 (with N=3, again in the summer in Amida, Syria/Turkey),
which was weaker than in AD 772 (unshifted), 
i.e. it happened in the strong activity decline.\footnote{See Sect. 3.2 and 3.3 for the dating of those aurorae.}
The strong increase in $^{14}$C around AD 774/5 (unshifted) lasted 3-4 yr in the different data sets.

From the minimum in about AD 774 (shifted) to the next previously established
minimum in about AD 784 (shifted), there may be one or even two short weak cycles --
like the lost cycle at the end of the 18th century (Usoskin et al. 2001), 
which was only 5.2 to 6.7 yr long (see footnote 1).
There are neither any sunspots nor any aurorae from AD 774 to 785, i.e. very low activity (high $^{14}$C level).
Afterwards, we then have a cycle from about AD 784 to AD 792 or 794 with 
significant radiocarbon amplitude and almost average length.
The activity (seen in radiocarbon and aurorae) continues to increase more or less continuously until AD 844,
to reach a similar $^{14}$C level as before AD 775.

We should now check whether the resulting Schwabe cycles could be different
without considering radiocarbon (as Schove 1955 and U13).
With our critically reviewed aurorae,
we see clusters of auroral activity 
in AD 743-746, 757-773, 793-796, 804-807, and 817-819,
all of which include at least one aurora 
with N$\ge 3$ and sometimes additional other aurorae (see Figs. 1 and 2).
The aurora cluster from AD 757 to 773 may well include two activity maxima 
of two separate Schwabe cycles, roughly AD 761 and 770. 
The years AD 734-735 and 786 also show at least two aurorae within one or two years,
but aurorae with N=1 or 2 only.
From the earliest to the latest {\em auroral} maximum, AD 734-735 to 817-819, 
one would expect nine Schwabe cycle maxima (cycles with normal length on average).
Given the clusters of auroral activity found above,
we would expect one more maximum between the clusters of AD 743-746 and 757-773,
i.e. around 751/2, and also one more maximum between the clusters of AD 757-773 and 786,
i.e. around AD 779/780.

Indeed, Schove (1955) has a weak maximum in AD 754, see Fig. 2;
such a maximum fits well in his list, because he has the next maximum in AD 765;
while we do not have access to the exact and full aurora list used by Schove (1955, 1984),
he probably concluded on a strong certain maximum in AD 765 partly using
the aurora in Amida dated to AD 766 (Schove 1984), but dated to AD 772 by us
(see Sect. 3.2 and 3.3);
the next three maxima ({\em uncertain} in Schove 1955) are also shifted by a few
years compared to our maxima (from aurorae and radiocarbon).
Instead of four cycles in Schove (1955) from AD 730 to 770, 
we have three cycles from about AD 733 to about AD 766.  

We could place one activity maximum between the aurora cluster
from AD 767-773 and the one in AD 786, and still would have average cycle length for all cycles.
Instead, we can also place two extra cycles with about 5 yr length each (see above and small arrow in Fig. 1).
After AD 784, we roughly agree with the reconstruction by Schove (1955)
about the next few cycles, all with normal length.

We still have to compare our results to U13:
U13 could have used aurorae together with $^{14}$C with high time resolution from M12 and/or U13,
but used aurorae only like Schove (1955). 
U13 claimed to have found
{\em a distinct cluster of aurorae between AD 770 and AD 776} suggesting 
{\em a high solar activity level around AD 775} plus {\em the next nearest
observations around AD 765-767 and AD 786, this suggests a 11-year cyclicity.} (U13). 
Let us briefly consider their events; even though U13 did not give individual details about 
all of their events, we can identify them quite certainly given the citations in U13 
(see Sect. 3.2 and 3.4 for details): \\
The presumable aurora in AD 775 (Dec, China: {\em above the moon}) 
was an event in AD 776 Jan 12/13 and most probable not an aurora;
the AD 773 Aug 9 observation in China ({\em 17-19h ... three bands of vapour}
was before the end of civil twilight (19:16h), hence not an aurora;
the {\em red cross} presumable for AD 773/4 (or 776) was a halo display in AD 776;
the {\em inflamed shields} was a halo display in AD 776 during day-time;
the {\em riders on white horses} presumably in AD 773 was in AD 774,
namely the frequent {\em dioscuri motive} as another form of narrating mock suns;
the remaining only possibly true aurorae among the events listed in U13 for the AD 770s are
one report from Ireland (AD 772, {\em  The assembly of the hand-clapping ... called fire from heaven})
and the two Chinese reports for AD 770 June and July ({\em white vapour appeared in the NW})
(maybe a duplication),
where neither colour nor night-time are mentioned. 
To conclude from {\em a distinct cluster of aurorae between AD 770 and AD 776} (U13) to
{\em a high solar activity level around AD 775} (U13) is possible only when giving a
higher weight to the events in (and around) AD 775; the only event listed for AD 775 in U13
happened in Jan 776 and was very likely not an aurora, 
and the events listed in U13 for AD 773/4 and 776 were halos.  \\
The presumable aurora cluster AD 765-767 (according to U13 citing Link 1962) consists of the following reports
from Europe: AD 765 Jan 1 in England ({\em fiery strokes}) and
AD 765 in Ireland ({\em A horrible and wonderful sign was seen in the stars at night}),
the former is a likely true aurora, but the unspecific Irish report may point to the same event
as the English report;
with {\em the next nearest observations at AD 767}, U13 may mean any of the following
three reports from the Far East: 
AD 767 Oct 8 in China ({\em At night white mist arose in the NW}),
indeed a possible aurora (but white, i.e. weak), 
AD 767 Aug 25 in China ({\em 17-19h ... white vapour spreading over the sky}, i.e. during
civil twilight, which ended at 18:56h local time for this place and time, hence not an aurora),
and AD 767 in Korea ({\em three stars drop down the garden of the Palace} indicating meteors).
Hence, the alleged aurora cluster and solar activity maximum in AD 765-767 in U13 is based on
only two possible aurorae on AD 765 Jan 1 in England and AD 767 Oct 8 in China. \\
The aurora cluster given in U13 for AD 786 is based on two likely true aurorae
reported for AD 786 Dec (one in China, one in Europe citing Link 1962) and one dubious report
(AD 786, Ireland: {\em A horrible vision in Cluain Moccu Nois, and great penance done 
throughout Ireland}).\footnote{U13 
wrote {\em In a new survey of occidental chronicles ... Further new Irish observations were
dated AD 765 and 786 and are found in the Annals of Ulster}, which probably means the two Irish reports
cited above; the event in AD 786 ({\em penance ... throughout Ireland}) was previously 
also considered to be an aurora by Schove (1964).} \\
U13 quoted both Keimatsu (1973) and Yau et al. (1995)\footnote{The Xu et al. (2000) book is
quoted by U13 only for its list of guest stars (novae), but not for aurorae; some of the events
still considered aurorae in U13 were already excluded by Xu et al. (2000).} 
for {\em credible observations from ... China} and then list all events from
Keimatsu (1973) for that period,\footnote{including events given as doubtful already in Keimatsu (1973)}
U13 do not discuss why they included some events which were already excluded by Yau et al. (1995).
On the other hand, a number of additional aurora reports for the AD 760s, 770s, and 780s, 
most of which published in astronomical literature, was not considered in U13.

Using published lists of possible aurorae only without a critical review of each event
-- or studying time spans which are too short -- can be misleading, see Fig. 2.
A major difference between the activity maxima as given in Schove (1955) for the whole 8th century 
and U13 for AD 765-786 on the one hand and our results above are as follows:
Both Schove (1955) and U13 give an activity maximum at AD 765 (AD 765-767 in U13),
while we found a maximum around AD 762 with three strong and credible aurorae reports 
from China in AD 762.
Also, both Schove (1955) and U13 give an activity maximum at around AD 775,
but based mainly on halo reports, while we found
auroral activity until AD 773, but then no aurorae az all until at least AD 786.

Given that there are several low- to mid-latitude aurora reports, e.g. Chinese and Arabic reports,
the Turkish/Syrian aurora sightings for the early 770s cannot be considered as exceptionally strong aurorae.
We see no additional $^{14}$C increases around any of the strong aurorae 
in our study period (Sect. 3.3).

We do not only agree with the statement by Cliver et al. (2014) that 
{\em the evidence for a grouping of strong auroral activity ca. 775 is weakened},
but -- even more -- we show that there is no evidence for
any report about likely true aurorae in the mid-770s (after AD 773).

The events listed in Sect. 3.2 and 3.4 also do not show any evidence for a global airglow event as
a possible indication for the nearby supernova or a Galactic gamma-ray burst around AD 775
(such an airglow was suggested by Pavlov et al. 2013).

\acknowledgements 
We thank F. Miyake for $^{14}$C data and J. Beer for $^{10}$Be in electronic form.
We retrieved the $^{14}$C Intcal09 data from www.radiocarbon.org 
and $^{10}$Be Dome Fuji data from ftp.nc dc.noaa.gov/pub/data/paleo/icecore/antarctica/domefuji/domefuji- 10be2008.txt
We are also grateful to M. Csikszentmihalyi and J. Chapman (UC Berkeley) for clarifying 
the date in the Chinese aurora observation dated here to AD 776 Jan (instead of 775 Dec)
and some additional translations of Chinese texts.
We also acknowledge N. Takahashi (U Tokyo) for kindly providing us with the papers by M. Keimatsu
and T. Posch (U Vienna) for accessing the book of Pilgram, which appeared
in 1788 in Vienna, which we could consult in their university observatory library.
We would like to thank R. von Berlepsch (AIP Potsdam) for providing us with a copy of the paper by Sch\"oning (1760),
whose relevant parts were then translated from old Danish by S. Buder (U Jena).
We acknowledge S. Daub, D. Luge, and P.A. Neuendorf (U Jena) for help with translations from Latin.
We would like to thank Dr. A. Harrak (U Toronto) for providing an English translation of the Syriac
text of the Chronicle of Zuqnin plus additional information and drawings of the two aurorae in AD 772/3,
which are located as original autograph in the Vatican library.
We obtained the data on moon phases from F. Espenak, NASA/GSFC, available
on eclipse.gsfc.nasa.gov/eclipse.html.
We also thank the Institut f\"ur Geschichte der Ara\-bisch-Islamischen Wissenschaften, U Frankfurt,
where we could consult the report about the Arabic sunspot observation in AD 840;
and we thank P. Kunitzsch (LMU Munich) for advise on the transliteration of Arabic names and words.
We consulted the Silverman aurora catalog on nssdcftp.gsfc.nasa.gov/miscellaneous/aurora,
we consulted the Ulster Chronicle (Mac Airt \& Mac Niocaill 1983) in its online Corpus of Electronic Texts Edition
on www.ucc.ie/celt/published/T100001A from the University College Cork, Ireland (P. Bambury \& S. Beechinor).
Monumenta Germania Historia (MGH) was consulted at their online version eMGH.
We thank T. Honegger (U Jena) for advice on the meaning of {\em \ae fter} in medieval English.
We acknowledge T. Zehe and F. Gie\ss ler (U Jena) for the calculation of the geomagnetic latitudes for Table 1.
We thank M. Korte (GFZ Potsdam) for advice on the geomagnetic field and for 
providing the coordinates of reconstructed pole positions in electronic form.
We had the opportunity to present this work at colloquia talks in Vienna, Munich, Potsdam,
and Napoli as well as conferences on the history of astronomy and physics, so that we would also like
to thank the organizers and attendees for useful discussion, in particular R. Arlt.
Especially, we would like to strongly acknowledge the Thuringian University and State Library 
(ThULB at U Jena), where we could find almost all the new and old literature used here,
and where we could also consult medieval hand-written manuscripts.
We thank D. Keeley for English language editing.
We would also like to acknowledge an anonymous referee for his comments and encouragment.

\end{document}